\documentclass[twocolumn]{aastex631}

\usepackage{xspace}
\usepackage{tikz}
\usepackage{float}

\newcommand{\swiftj}{Swift~J1727.8$–$1613\xspace}
\newcommand{\eighteentwenty}{MAXI~J1820+070\xspace}

\newcommand{\gx}{GX~339$-$4\xspace}

\newcommand{\distance}{5.5\,kpc\xspace}

\received{January 21, 2026}
\revised{April 14, 2026}
\accepted{April 15, 2026}

\graphicspath{{./}{figs/}}

\begin{document}

\title{Systematic assessment of disk truncation in the black hole X-ray binary \swiftj using NICER}

\newcommand{\cfa}{Center for Astrophysics $\vert$ Harvard \& Smithsonian, 60 Garden Street, Cambridge, MA 02138, USA}
\newcommand{\remeis}{Dr. Karl Remeis-Observatory and Erlangen Centre for Astroparticle Physics, Friedrich-Alexander-Universit\"at Erlangen-N\"urnberg, Sternwartstr.~7, 96049 Bamberg, Germany}
\newcommand{\massit}{MIT Kavli Institute for Astrophysics and Space Research, Cambridge, MA, 02139, USA}
\newcommand{\gsfc}{NASA Goddard Space Flight Center, Astrophysics Science Division, 8800 Greenbelt Road, Greenbelt, MD 20771, USA}
\newcommand{\amsterdam}{Anton Pannekoek Institute for Astronomy, University of Amsterdam, Science Park 904, NL-1098 XH, Amsterdam, The Netherlands}
\newcommand{\czechacademysciences}{Astronomical Institute of the Czech Academy of Sciences, Bo\v{c}n\'i II 1401/1, 14100 Praha 4, Czech Republic}

\correspondingauthor{Ole~K\"onig}
\email{ole.koenig@cfa.harvard.edu}

\author[0000-0001-8670-4575]{Ole~K\"onig} \affiliation{\cfa}

\author[0000-0002-5872-6061]{James~F.~Steiner}\affiliation{\cfa}

\author[0009-0005-6609-5852]{Niek~Bollemeijer}\affiliation{\amsterdam}

\author[0000-0002-8908-759X]{Riley~M.~T.~Connors}
\affiliation{Department of Physics, Villanova University, 800 E. Lancaster Avenue, Villanova, PA 19085, USA}

\author[0000-0003-4583-9048]{Thomas~Dauser} \affiliation{\remeis}

\author[0000-0003-0079-1239]{Michal~Dov\v{c}iak}
\affiliation{\czechacademysciences}

\author[0009-0009-2549-1161]{Ningyue~Fan}
\affiliation{Department of Physics, Stanford University, Stanford, CA 94305, USA}
\affiliation{Kavli Institute for Particle Astrophysics and Cosmology, Stanford University, Stanford, CA 94305, USA}

\author[0000-0003-3828-2448]{Javier~A.~Garc\'ia} \affiliation{\gsfc}

\author[0000-0002-6288-4791]{David~Horn} \affiliation{\remeis}

\author[0000-0002-5311-9078]{Adam~Ingram}
\affiliation{School of Mathematics, Statistics and Physics, Newcastle University, Herschel Building, Newcastle upon Tyne NE1 7RU, UK}


\author[0000-0002-2235-3347]{Matteo~Lucchini} \affiliation{\amsterdam}

\author[0000-0003-4216-7936]{Guglielmo~Mastroserio}
\affiliation{Scuola Universitaria Superiore IUSS Pavia, Palazzo del Broletto, Piazza della Vittoria 15, I-27100 Pavia, Italy}

\author[0009-0003-1304-9014]{Cal~Miller} \affiliation{\cfa}

\author[0000-0002-9633-9193]{Edward~Nathan} \affiliation{\gsfc}

\author[0000-0001-6923-1315]{Michael~A.~Nowak}
\affiliation{Department of Physics, Washington University in St. Louis, Campus Box 1105, One Brookings Drive, St. Louis, MO 63130-4899, USA}

\author[0000-0002-4656-6881]{Katja~Pottschmidt$^\dag$} \affiliation{\gsfc}

\author[0000-0003-4815-0481]{Ron~Remillard} \affiliation{\massit}

\author[0009-0001-0703-2000]{Yujia~Song}
\affiliation{Key Laboratory for Computational Astrophysics, National Astronomical Observatories, Chinese Academy of Sciences, Datun Road A20, Beijing 100012, China}
\affiliation{School of Astronomy and Space Sciences, University of Chinese Academy of Sciences, Datun Road A20, Beijing 100049, China}

\author[0000-0003-2931-0742]{Ji\v{r}\'i~Svoboda}
\affiliation{\czechacademysciences}

\author[0000-0003-0070-9872]{Michiel~van~der~Klis}\affiliation{\amsterdam}

\author[0000-0002-6159-5883]{Santiago~Ubach}\affiliation{Departament de F\'isica \& CERES-IEEC, Universitat Aut\`onoma de Barcelona, Bellaterra, Spain}

\author[0000-0003-2065-5410]{Jörn~Wilms} \affiliation{\remeis}

\author[0000-0002-2268-9318]{Yuexin~Zhang} \affiliation{\cfa}
\affiliation{Kapteyn Astronomical Institute, University of Groningen, P.O.\ BOX 800, 9700 AV Groningen, The Netherlands}


\begin{abstract}
The 2023/24 NICER monitoring campaign of the 7~Crab bright black hole X-ray binary \swiftj covered the outburst in almost all accretion states. High-quality data are available in the high-Eddington-fraction hard-intermediate state, hard-to-soft transition, the soft state, and the poorly studied back-transition to the dim hard state, making it an ideal dataset to compare the accretion flow at vastly different accretion rates. We apply disk continuum fitting techniques to investigate the evolution of the inner disk radius throughout the outburst. Taking a temperature-dependent color-correction factor into account, we see evolution of the disk inner radius by a factor of a few comparing the hard states to the thermal/soft state. We tentatively detect an onset of disk truncation in the soft-to-hard transition, right after the source leaves the soft state. After accounting for model systematics, we find the disk to be more truncated in the high-luminosity bright hard state compared to the low-luminosity dim hard state.
\end{abstract}



\section{Introduction} \label{sec:intro}

The two most dominant emission components of black hole X-ray binaries in outburst are thermal emission from an accretion disk and Comptonization of these soft seed photons in an optically thin plasma often called the corona \citep{ShakuraSunyaev1973a}. 
The standard outburst paradigm of transient low-mass X-ray binary black holes (LMXB BHs) predicts that there are significant changes in the inner radius of the accretion disk between different states \citep{RemillardMcClintock2006a}. 
In quiescence, the accretion disk is thought to be highly truncated at ${\sim}10^3$--$10^4\,r_g$ \citep[e.g.,][]{NarayanMcClintockYi1996a,McClintock2003a}, where $r_g=GM/c^2$ is the gravitational radius.
As the source increases in luminosity, it passes through a dim and bright hard state where the major contribution to the energy flux is from the corona.
It is unclear whether the disk truncation radius in these states has already decreased to the innermost stable circular orbit (ISCO) or not.
After the source has transitioned to the disk-dominated soft state, the disk is generally considered to extend to the ISCO, both from the observational \citep[e.g.,][]{TanakaLewin1995a,KubotaMakishima2004a,GierlinskiDone2004a,Gou2009a,Steiner2010a} and theoretical standpoint \citep[e.g.,][]{Penna2010a}. 
When the source transitions back into a dim hard state at a lower luminosity of one to a few percent of the Eddington luminosity \citep{Maccarone2003a,VahdatMotlagh2019a}, the disk temperature decreases and the corona again becomes more prominent. This soft-to-hard transition happens on the order of days to weeks, and the onset of truncation as the source goes back into quiescence is not well established.

Significant progress has been made in understanding the inner truncation radius of the accretion disk between the extreme points of quiescence and the soft state. However, finding a conclusive physical interpretation is difficult for several reasons. 
First, physical models of accreting black holes are highly dependent on the assumption of the geometry and the emissivity profile of the accretion flow, particularly, of the corona.
As a result, in the bright hard state, some studies measure a highly truncated disk with an inner hot flow \citep[e.g.,][]{Done2007a,Plant2015a,Marcel2019a,Mahmoud2019a,ZdziarskiDeMarco2020a,Zdziarski2021a,Kawamura2022a}, while other analyses derive a truncation radius consistent with the ISCO \citep{Miller2006a,Reis2008a,Tomsick2008a,Reynolds2010a,ReynoldsMiller2013a,Garcia2015a,Kara2019a,Sridhar2020a,Connors2022a}. 

A second complication is that spectral modeling requires describing complex astrophysical processes with a chosen set of parameters that should be interpretable. 
The ``chosen set of parameters'' refers to employing parameterized spectral models that do not explicitly calculate all known physics but use simplifying approximations, and, therefore, necessarily, choices that are open to criticism.
Understanding the physical meaning and implicit biases (``systematics'') of these choices is not trivial.
Particularly relevant for our paper is the effect of spectral hardening. This effect arises from the radiative transfer through the disk atmosphere that modifies the radiation of the underlying thermal disk from a blackbody at the temperature of the disk midplane \citep{Madej1974a}.  
Specifically, the opacity in the disk atmosphere is predominantly electron scattering, which means that the observer sees deeper into the disk, where it is hotter, thereby hardening the observed spectrum (for more details, see appendices of \citealt{DavisDoneBlaes2006a} and \citealt{SalvesenMiller2021a}, and references therein). We note that this interior temperature is further increased from the blanketing effect of the electron-scattering atmosphere \citep{FeltenRees1972a,LondonTaamHoward1986a}. 
These effects of radiative transfer within the disk atmosphere can be parameterized by a color-correction factor \citep{ShimuraTakahara1995a}, $f_\mathrm{col}=T_\mathrm{col}/T_\mathrm{eff}\sim 1.3$--2.0 (see, however, \citealt{Salvesen2013a,SalvesenMiller2021a}) that connects the emergent color temperature of the disk, $T_\mathrm{col}$, to the effective temperature, $T_\mathrm{eff}$, of the black body that produces the same luminosity.
Correctly accounting for spectral hardening is important when interpreting physical model parameters. For instance, it is known that there is a strong degeneracy of $f_\mathrm{col}$ with the black hole spin (a higher spin can be compensated by a lower $f_\mathrm{col}$; e.g., Table~3 of \citealt{Li2005a}). 

Third, it is challenging for a model to explain all observational data at the same time. As in the ``blind man touching the elephant'' analogy, different diagnostics of the data (reflection fitting, continuum spectrum, accretion rate variability, polarimetry, radio jet observations) serve to gauge different physical regions of the accretion flow. Depending on the investigation, the effects associated with the ``corona'' under investigation may actually refer to subtly different entities, probing different locations of the accretion flow.
One result of this is that some studies have proposed geometries containing multiple coronae \citep[e.g.,][]{Nowak2011a,Bellavita2022a,GarciaFederico2022a,Lucchini2023a,Alabarta2025a}.

To assess these questions about the accretion flow, in this paper, we investigate the inner truncation radius of the accretion disk in the LMXB BH \swiftj using the continuum fitting method.
Disk continuum fitting relies on an accurate understanding of the temperature gradient in the geometrically thin, optically thick accretion disk. From standard disk theory \citep{ShakuraSunyaev1973a}, as implemented for the \texttt{diskbb} model, the temperature profile is $T_\mathrm{eff}(r)\propto r^{-3/4}$ \citep[][p.~755]{Mitsuda1984a}. The innermost part of the disk defines the location of the peak of the multi-temperature thermal disk spectrum. 
In continuum fitting, one fits for this observed color temperature at the inner disk edge (or slightly outwards in the case of \texttt{ezdiskbb}) and derives the effective temperature using $f_\mathrm{col}$. One then empirically determines the disk luminosity from the flux and uses that luminosity and the effective temperature to determine the inner radius according to (see \citealt{ShimuraTakahara1995a} and \citealt{Zimmerman2005a}, Eq.~8, for the proportionality factor)
\begin{equation}
    \label{eq:L_disk}
    L_\mathrm{disk} \propto r_\mathrm{in}^2 T_\mathrm{eff}^4 = r_\mathrm{in}^2 (T_\mathrm{col}/f_\mathrm{col})^4 \quad .
\end{equation}
In practical terms, one fits the disk normalization, $K$, to relate the disk luminosity to an inner radius using a distance, $D$, and inclination estimate, $i$, according to \citep[][Eq.~7]{Zimmerman2005a}
\begin{equation}
    \label{eq:norm_radius}
    K = \frac{1}{f_\mathrm{col}^4} \left ( \frac{r_\mathrm{in}}{D}\right )^2 \cos i \quad .
\end{equation}
As the temperature of the inner accretion disk edge in LMXB BHs is typically $\lesssim1$\,keV, disk continuum fitting requires coverage of soft X-ray energies, which can be measured with high fidelity using the \textit{Neutron Star Interior Composition Explorer} (NICER).

\swiftj (abbreviated as J1727 from now on) is a transient LMXB BH, presumably with an early K-type companion star (\citealt{MataSanchez2024a,MataSanchez2025a}, but see also \citealt{Burridge_2025a}), that went into a ${\sim}7.6$\,Crab bright outburst beginning in August 2023 \citep{Lipunov2023a_GCN,Page2023a_GCN,Negoro2023a_ATel,Palmer2023a_ATel}. The most recent distance estimate is $5.5^{+1.4}_{-1.1}$\,kpc, inferred from the radial velocity curve of radio observations, although \citet{Burridge_2025a} note that this value may be smaller if the donor star has a stripped envelope.
J1727 has been of large interest to the black hole binary community due to its bright atypical hard state that exhibited periods of intense flaring \citep{Liao2025a}, and its comparatively faint soft state (similar behavior has been seen in GRS~1758$-$258, see \citealt{Pottschmidt2006a,Pottschmidt2008a}). The source has been extensively studied in the spectral \citep[e.g.,][]{Peng2024a,Svoboda2024a} and timing \citep{Mereminskiy2024a,Nandi2024a,Yu2024a,ZhuWang2024a,Bollemeijer2025a,Brigitte2025a} domains. A striking result from X-ray polarization measurements \citep{Veledina2023a,Ingram2024a,Svoboda2024a,Podgorny2024a} is the similarity of the bright and dim hard states of J1727, with a similar X-ray polarization degree (despite the very large difference in luminosity) and an angle that is parallel to the radio jet in both cases \citep{Vrtilek_ATEL16230,Bright_ATel16228,MillerJones_2023ATel16211,Wood2024a}, indicating a corona that is primarily extended along the plane of the accretion disk. 
The polarimetry measurements indicate an intermediate inclination in the range 30--$50^\circ$ \citep{Svoboda2024a}, consistent with an upper limit of ${<}69^\circ$ inferred from radio radial velocity measurements \citep{Burridge_2025a}.
We fix the inclination to an intermediate value of $45^\circ$ in this paper. 
Dynamical measurements of J1727 establish a firm lower limit on the compact object mass and confirm that it is a black hole \citep{MataSanchez2025a}. However, a precise value for the black hole's mass has not been published, yet. Thus, in line with previous studies, we assume a canonical black hole mass of $10\,M_\odot$.

This paper is structured as follows. Section~\ref{sec:data_reduction} describes the extraction and orbit-filtering of the NICER spectra. In Sect.~\ref{sec:results}, we fit each spectrum with a continuum model and classify the observations into spectral states. We discuss physical implications on disk truncation in Sect.~\ref{sec:discussion} and highlight the complexity in how model assumptions influence the interpretation. We summarize our findings on the disk truncation debate in Sect.~\ref{sec:conclusions}, highlighting tentative truncation trends in the hard states.

\section{Data reduction and selection}
\label{sec:data_reduction}

We process the NICER data with HEASOFT 6.34, NICERDAS 12, and CALDB xti20240206.
With \texttt{nicerl2}, we prepare a ``night'' dataset, selecting only orbit-night data with an undershoot range below $25\,\mathrm{s}^{-1}$. This conservative choice is motivated by our goal to measure the properties of the disk in the dim hard state, where the temperature and flux of the disk are low (a higher undershoot rate can lead to increased low-energy noise, which may otherwise bias our disk measurements).
For all data in this paper, we use the default overshoot threshold of ${<}30\,\mathrm{s}^{-1}$ to exclude high-energy background events.
We then use \texttt{maketime} to create good time intervals (GTIs) from the cleaned event file, requiring a minimum exposure of 100\,s. With \texttt{nicerl3-spect}, we then extract spectra for each GTI and use the default systematic errors calculated in the extraction pipeline (1.5\% across most calibrated energy channels).
In the orbit-night dataset, we obtain 469 spectra which we analyze with ISIS version 1.6.2-51 \citep{Houck2000a}. These data are distributed as 100\,ks of exposure in the hard-intermediate state (HIMS), 30\,ks in the hard-to-soft transition region, 42\,ks in the soft state, and 22\,ks in the back-transition and dim hard state.

In addition, the archive contains 76\,ks of exposure during orbit day, mainly distributed in the flaring period during the HIMS and throughout the back-transition into quiescence. Due to the light leak issue\footnote{\url{https://heasarc.gsfc.nasa.gov/docs/nicer/data\_analysis/nicer\_analysis\_tips.html\#lightleakincrease}} and the resulting noise in the low-energy passband (we filter for an undershoot range of 0--$500\,\mathrm{s}^{-1}$), we analyze and label these ``day'' observations separately throughout the whole paper. 
We emphasize that the inclusion of these day data is highly beneficial for constraining the parameter evolution across the outburst as it samples times of significant changes in the source where no orbit-night data is available, particularly during the back-transition into the dim hard state.

We fit the NICER data in the energy range 0.45--10\,keV with the SCORPEON background model.
In Appendix~\ref{sec:app:noisepeak}, we show that for J1727, a soft component impacts the energy range below 0.45\,keV that may or may not be physical. In this paper, we limit the energy range analyzed to above 0.45\,keV to avoid biases from this unknown component. 
In Appendix~\ref{app:sec:NICER_sensitivity}, we show that, with this threshold, we can constrain disk temperatures down to around 60\,eV, which is more than a factor of two smaller than the lowest temperature we fit in this dataset.

Spectra are re-binned with the optimal binning scheme of \citet{KaastraBleeker2016a} with a minimum number of 20 counts per bin, and fitted using the $\chi^2$-statistic. All uncertainties quoted are at the 90\% confidence level unless stated otherwise. 
We use a logarithmic energy grid between 0.01--1000\,keV with 1000 grid points, which is necessary for the Comptonization model \texttt{simpl} as the disk seed spectrum extends to energies below the NICER response.

\section{Results}
\label{sec:results}

\subsection{Description of the continuum model}
The goal of this paper is to test for a change in the inner disk radius by studying the thermal radiation of the accretion disk. We investigate large-scale relative changes throughout the whole outburst, and first deploy simple, well-understood disk models that have traditionally been used for such studies \citep[e.g.,][]{GierlinskiDone2004a}. Later (Sect.~\ref{sec:relativist_disk_models}), we show the results obtained with relativistic disk models that have only a minor impact on the resulting conclusions.

The \texttt{diskbb} model \citep{Mitsuda1984a,Makishima1986a} is the most widely used non-relativistic standard thin disk model. It assumes non-zero torque at the inner disk edge \citep[e.g.,][]{GierlinskiDone2004a}. The \texttt{diskbb} model itself, however, does not exhibit any radiating material interior to the disk, resulting in a discontinuity of the radial temperature gradient at the inner disk edge. A variation of this model is \texttt{ezdiskbb}, which imposes a zero-torque boundary condition at the inner edge \citep{Zimmerman2005a}. This model has no discontinuity, and the maximal color temperature, $T_\mathrm{max}$, is slightly outwards of the inner edge (e.g., Fig.~A1 of \citealt{Gierlinkski1999a}).
In practice, the main differences between \texttt{diskbb} and \texttt{ezdiskbb} are the absolute values of the temperature ($T_\mathrm{max}$ in \texttt{ezdiskbb} is smaller than $T_\mathrm{in}$ in \texttt{diskbb} for a given disk inner radius and accretion rate, see Eq.~5 of \citealt{Zimmerman2005a}), which results in higher values of the normalization in \texttt{diskbb}. For the overall trend-lines across the outburst, which we are interested in this paper, this choice has minor implications. 
Assuming a null hypothesis of no truncation, we argue that, unless there is intra-ISCO emission (see also Sect.~\ref{sec:model_systematics}), \texttt{ezdiskbb} is physically more realistic than \texttt{diskbb} \citep[see also][]{Paczynski2000a}, and we will deploy it as the main disk continuum model in this paper. 

For the Comptonization, we use \texttt{simpl} \citep{Steiner2009a}, which is an empirical, convolutional model that implements the Compton scattering of the disk seed photons into the corona. We allow for up and down-scattering of photons (\texttt{UpScOnly=0}). In this model, the scattered fraction ($f_\mathrm{sc}$) denotes the percentage of seed photons being Comptonized by the corona. 
Furthermore, we do not see strong reflection features in the NICER passband. Therefore, no detailed reflection modeling is applied for these data (see discussion on systematics in Sect.~\ref{sec:relativistic_reflection}), and we model the iron line region using the \texttt{laor} model \citep{Laor1991a}. 
We also include two narrow empirical Gaussian lines at $1.74\pm 0.05$\,keV and $2.20\pm 0.05$\,keV, allowing for positive and negative normalization, to mitigate uncalibrated features around the Si~K line and Au~M edge, respectively. 
The full baseline model in XSPEC terminology is 
\begin{equation}
    \label{eq:simplXezdiskbb}
    \texttt{tbfeo*(simpl\textasteriskcentered ezdiskbb+laor)+2$\times$gauss} \quad ,
\end{equation}
where \texttt{tbfeo} accounts for interstellar absorption \citep{Wilms2000a}.
LMXBs in outburst typically do not show significant changes in their intrinsic absorption as they go through an outburst. Consequently, many studies fix the equivalent hydrogen column density ($N_\mathrm{H}$) to a fixed value throughout the whole outburst \citep[e.g.,][]{Miller2009a,Garcia2019a}.
In Appendix~\ref{sec:app:abundances}, we fit all observations with free $N_\mathrm{H}$ and observe scatter around the mean by roughly $\pm 4\%$. We infer that the assumption of constant absorption is reasonable for J1727. In Appendix~\ref{sec:app:abundances}, we also discuss the usage of non-solar abundances ($A_\mathrm{O}=1.06$ and $A_\mathrm{Fe}=0.45$) that we adopt hereafter. These abundance values are plausibly physical, but alternatively could account for few-percent-scale errors in NICER's calibration at low energies. Our final absorption estimate is inferred from a simultaneous fit to 15 spectra sampling the HIMS, soft state, and dim hard state: $N_\mathrm{H}=0.2597\times 10^{22}\,\mathrm{cm}^{-2}$. This value is consistent with previous studies on this source \citep{HI4PI2016a,Svoboda2024a,Chatterjee2024a}.  

\begin{figure}
    \includegraphics[width=1\linewidth]{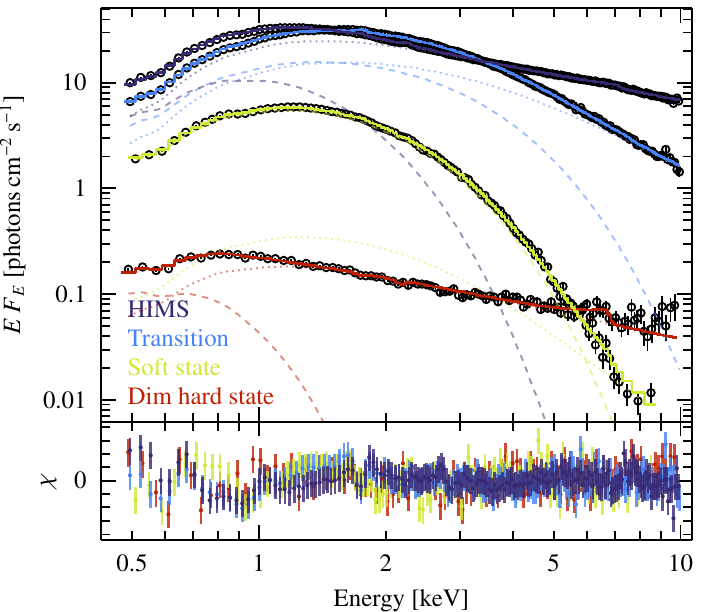}
    \caption{Example spectra of \swiftj at different stages of its outburst, modeled with \texttt{simpl\textasteriskcentered ezdiskbb} (Eq.~\ref{eq:simplXezdiskbb}). The Obs. IDs and $\chi^2_\mathrm{red.}$ in these representative HIMS, hard-to-soft transition, soft state, and dim hard state observations are, respectively, 6203980108 with 96.2/(175-9)=0.58, 6203980131 with 95.5/(157-9)=0.65, 6203980142 with 116.8/(110-9)=1.16, and 7708010112 with 113.3/(111-9)=1.11. Dashed and dotted lines show the disk and Comptonization components, respectively. The color scale is based on the MJDs of the observations as in Fig.~\ref{fig:q_diagram}.}
    \label{fig:spectra}
\end{figure}

Representative spectra for the different phases of the outburst can be found in Fig.~\ref{fig:spectra}.
We find that the model slightly overfits the earlier NICER monitoring data, when the source is hard. This may be an indication that the default 1.5\% systematic uncertainty from \texttt{nicerl3-spect} is overestimated in this data group.
When the source is soft, the model is slightly underfitting, which is mainly due to broad features below 2\,keV that we do not account for (this structure can be seen well in Appendix Fig.~\ref{fig:multifit_all}). Towards quiescence, it is likely that higher relative background (which is a modeled quantity, not directly measured), results in additional systematic uncertainty. 
Overall, we find that further optimization of the fit statistic has minor impact on the large-scale evolution of the parameters that we study in this paper (note the large differences in the spectra in Fig.~\ref{fig:spectra}). 

\subsection{Overall behavior of \swiftj in the outburst cycle of transient LMXB black holes}

\begin{figure}
    \centering
    \includegraphics[width=1\linewidth]{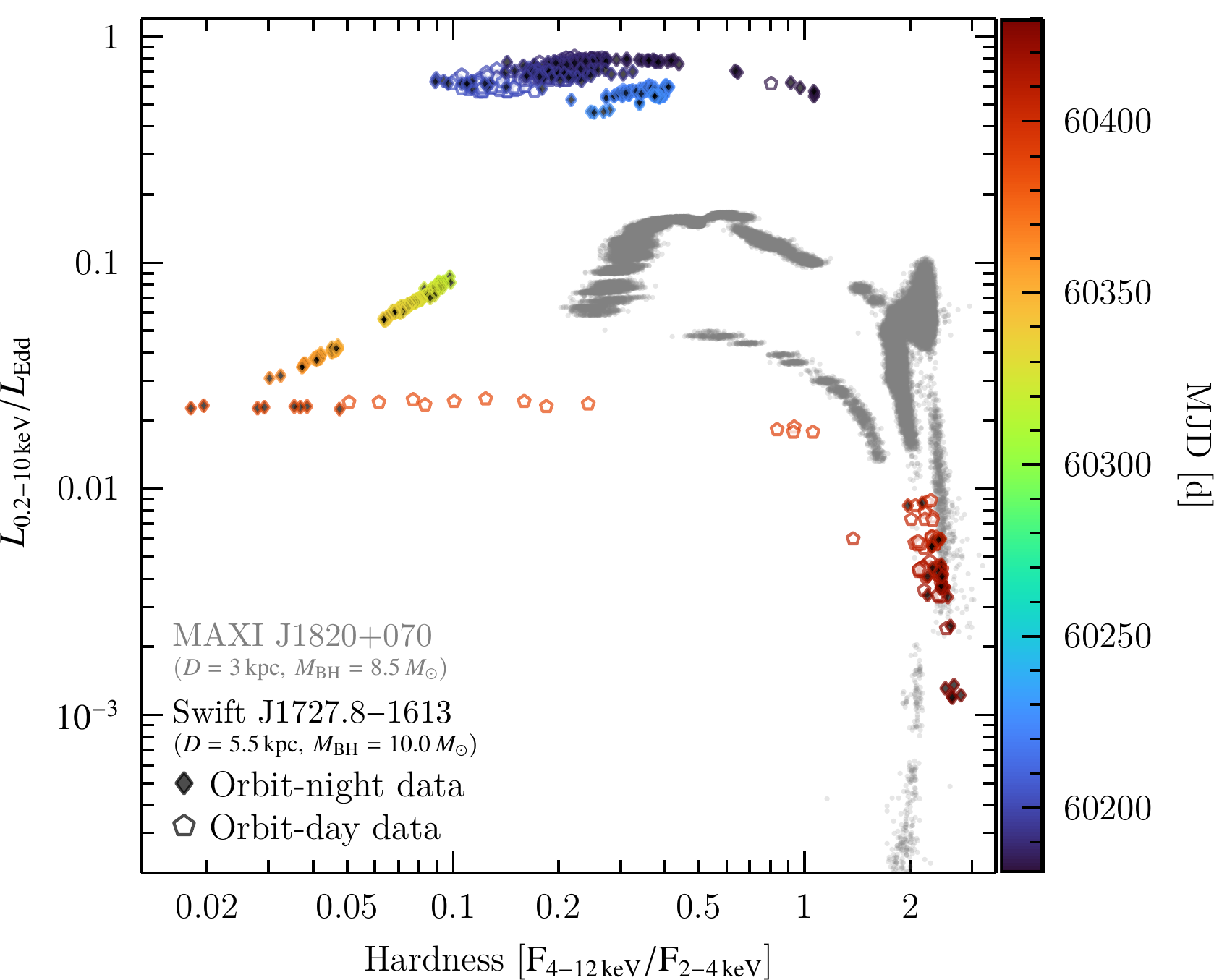}  
    \caption{Hardness-luminosity diagram of \swiftj. The soft and hard bands are 2-–4\,keV and 4–-12\,keV, respectively, and the hardness is derived from (unabsorbed) fluxes of phenomenological model fits. The 0.2--10\,keV Eddington luminosity fraction is derived from \texttt{simpl\textasteriskcentered ezdiskbb} model fits assuming a black hole of mass $10\,M_\odot$ and a distance of \distance. The point where we detect a tentative onset of disk truncation is on the horizontal orange line where the orbit-night data (diamonds) passes to the orbit-day data (open pentagons).
    NICER data of \eighteentwenty are calculated with the same procedure and shown for reference.}
    \label{fig:q_diagram}
\end{figure}

We first show the overall behavior of J1727 across its outburst in Fig.~\ref{fig:q_diagram}. Instead of a hardness-intensity diagram, we calculate the hardness directly from the (unabsorbed) model fluxes from our fits, thereby removing differences in Galactic absorption and effects from the effective area of the telescopes (see \citealt{Barillier2023a} and \citealt{Koenig2024a} for details). The intrinsic X-ray luminosity is calculated from the unabsorbed \texttt{simpl\textasteriskcentered ezdiskbb} model flux in the 0.2--10\,keV range, assuming a distance of \distance for J1727, and scaled to the Eddington luminosity assuming a fiducial BH mass of $10\,M_\odot$.

The hardness-luminosity diagram shows that J1727 traces the typical, roughly q-shaped outburst hysteresis of transient LMXB black holes. The NICER data sample multiple regions of this hysteresis, particularly the ``forward-transition'' on an upper branch and the ``backward-transition'' on a lower branch. There are some gaps in coverage, especially during the early rise at large hardness ratios and also at several intervals at small hardness.
We also compare J1727 to the NICER data of \eighteentwenty (J1820 from now on), explored via the same spectral analysis, in order to understand how J1727 behaved relative to another well-studied transient LMXB black hole.
Overall, we observe that J1727 reached higher luminosity than observed in the upper branch of the hardness-luminosity diagram in J1820 (Fig.~\ref{fig:q_diagram}, but note that the luminosity only includes the 0.2--10\,keV band, which can miss substantial flux during harder corona-dominated states compared to a bolometric luminosity), and also became significantly softer and fainter. Thus, our data are consistent with the hypothesis that the source was at or even above the Eddington limit on the upper branch \citep[e.g.,][]{2024Liu_arXiv}.

\subsection{State classification}
\label{sec:state_classification}
The hardness-luminosity diagram in Fig.~\ref{fig:q_diagram} shows that the source continuously softened on the upper branch of the q-diagram during the first phase of the outburst. Following \citet{2024Liu_arXiv}, we distinguish the first phase into a bright hard state (MJD 60181--60186), and a hard-intermediate state (MJD 60186--60198).
We adopt the same classification in order to facilitate easier comparison to the Insight-HXMT studies (e.g., \citealt{Ma2025a,Yang2024a,Bollemeijer2025a}), although we emphasize that as there is no clear line between the bright hard and hard intermediate states, this has minor implications on the interpretation.
Beginning on MJD 60198, J1727 started undergoing intense X-ray flaring \citep{Yu2024a,Liao2025a}, accompanied by several radio flares \citep{Hughes2025a} and ballistic ejections of jet knots \citep[][e.g., on MJD 60206]{Wood2025a}. We term this atypical phase the ``flare state.''
Several studies have proposed a hard-to-soft state transition at around MJD 60223 when another radio flare occurred \citep{MillerJones2023_ATel16271,Hughes2025a} and the X-ray variability properties changed \citep{Bollemeijer2023ATel16273,Yu2024a}. We therefore label the data group between MJD 60222--60227 as ``transition.''
Following the hard-to-soft transition, NICER did not observe J1727 for three months due to Sun constraints. In the data group MJD 60317--60383.6, the disk-dominated spectrum can be confidently assigned to the soft state with constantly decreasing flux.
The source then sharply turns to the right in the hardness-luminosity diagram (MJD 60383.6--60392), which we associate with the soft-to-hard transition (consistent with \citealt{Hughes2025a}). The data group taken afterwards (MJD 60403 until roughly 60425) can be associated with the dim hard state. There are some more NICER observations after our dim hard state group (until MJD 60462), however, the source here was too faint to produce robust constraints of the spectral parameters. 

\subsection{Spectral parameter evolution}
\label{sec:evolution}

\begin{figure}
    \centering
    \includegraphics[width=1\linewidth]{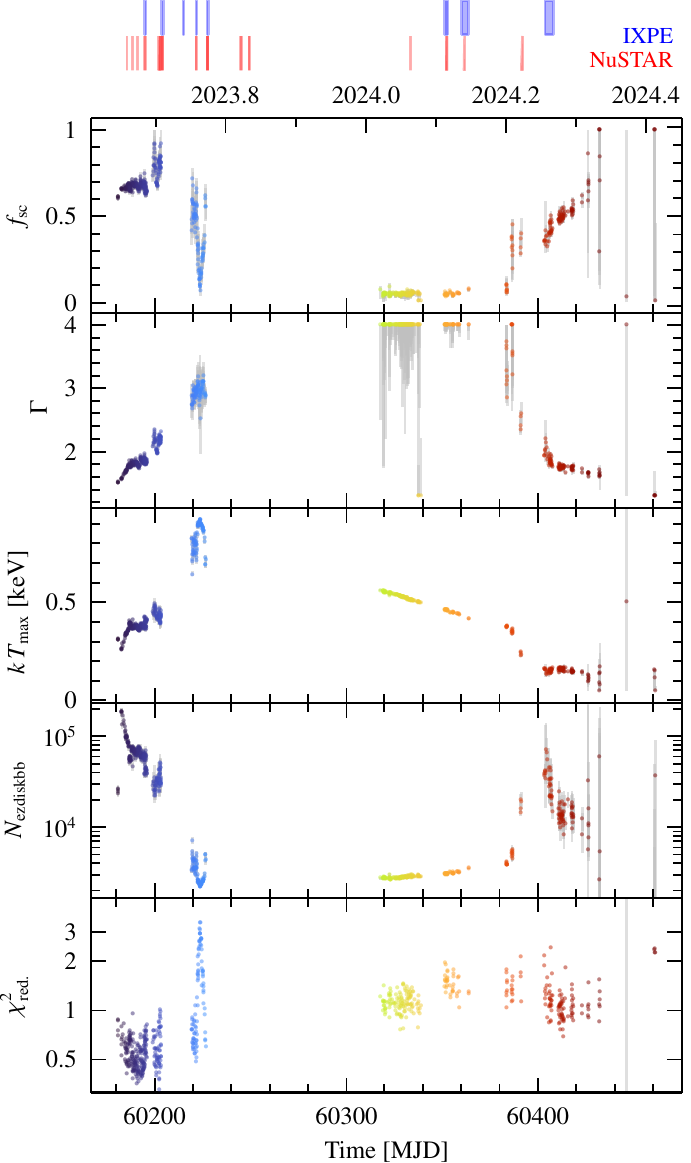}
    \caption{Spectral parameter evolution of \swiftj using the \texttt{simpl\textasteriskcentered ezdiskbb} model. Color denotes times with the same scale as in Fig.~\ref{fig:q_diagram}. Gray error bars are at the 90\% confidence level. To facilitate comparison to other studies of J1727, we show the observation times of IXPE and NuSTAR as blue and red bars on the top.}
    \label{fig:spectral_parameter_evolution}
\end{figure}

With the continuum model presented in Eq.~\ref{eq:simplXezdiskbb}, we obtain acceptable fits ($\chi^2_\mathrm{red.}\lesssim 2$) throughout most of the dataset, shown in the bottom panel of Fig.~\ref{fig:spectral_parameter_evolution}. The only exception is the hard-to-soft transition around MJD 60223 when our Comptonized standard thin disk model does not adequately describe the data. The spike in $\chi^2_\mathrm{red.}$ lasts for roughly four days. We briefly discuss this period in Sect.~\ref{sec:hard-to-soft-transition}. 

We find significant parameter changes across the roughly 260 days of the outburst, presented in Fig.~\ref{fig:spectral_parameter_evolution}.
The scattered fraction ($f_\mathrm{sc}$) gives the portion of up-scattered photons from the seed disk spectrum. As expected, this parameter is high in the hard state (many photons are Compton scattered) and low in the disk-dominated soft state. Likewise, $\Gamma$ shows a clear trend to higher values as the source softens. 
The flux, photon index, and disk temperature rise strongly at the beginning of the NICER observing window. After a few days, these changes become more gradual. 
In the soft state, it pegs at the maximal value of 4, which is expected given the minor contribution of a power-law component in this state (which is also reflected in the large error bars), and then decreases again as the source transitions through the dim hard state into quiescence. The disk's peak temperature ($kT_\mathrm{max}$) increases throughout the first phase of the outburst, from roughly 0.2\,keV at the beginning of the monitoring, with a maximum around the hard-to-soft transition (around MJD 60223), and then decreasing gradually with a shorter-timescale jump in the soft-to-hard transition at around MJD 60386.
Finally, there is significant evolution in the disk normalization, potentially pointing to changes in the inner radius, which we will interpret in Sect.~\ref{sec:radius_temperature_plane}. Here, we only comment on the non-intuitive fact that the disk normalization is higher at the beginning of the monitoring data compared to the soft state, even though Comptonization dominates the total flux in the hard states. This property can be explained by the strong temperature dependence of the disk luminosity, and the fact that the source at the beginning of the monitoring is a factor of roughly 10 higher in luminosity compared to the soft state, while the accretion disk is roughly 0.15\,keV cooler. As $L\propto r_\mathrm{in}^2 T^4\propto K T^4$ (Eq.~\ref{eq:L_disk} and \ref{eq:norm_radius}), a higher temperature in the soft state requires a lower normalization at a fixed radius.

\subsection{Luminosity evolution}
\label{sec:luminosity}

\begin{figure}
    \centering
    \includegraphics[width=1\linewidth]{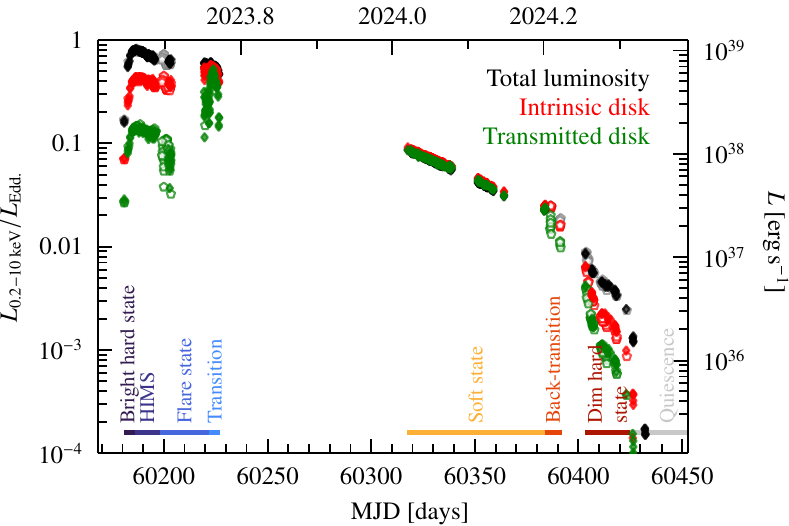}
    \caption{X-ray luminosity evolution of \swiftj, assuming a black hole mass of $10\,M_\odot$ and distance of \distance. The total model luminosity (Eq.~\ref{eq:simplXezdiskbb} without the absorption of \texttt{tbfeo}) is shown in black. The seed disk luminosity (\texttt{ezdiskbb}) is shown in red. The transmitted disk ($\texttt{ezdiskbb}\cdot (1-f_\mathrm{sc})$) is shown in green. Diamonds and open pentagon denote orbit night and day data, respectively.}
    \label{fig:luminosity_evolution}
\end{figure}

To quantify the relative contribution of the disk, we calculate the disk luminosity and show its time evolution across the outburst in Fig.~\ref{fig:luminosity_evolution}. The total (unabsorbed) disk luminosity in the 0.2--10\,keV band (i.e., non-bolometric) is calculated from Eq.~\ref{eq:simplXezdiskbb} without \texttt{tbfeo}. The intrinsic disk emission, that is, the thermal seed photon distribution, is calculated by evaluating only the \texttt{ezdiskbb} component within \texttt{simpl}. The transmitted disk, that is, the thermal radiation that is un-scattered and visible to the observer, is calculated by $\texttt{ezdiskbb}\cdot (1-f_\mathrm{sc})$ (see also \citealt{Steiner2009a}).

\begin{figure}
    \centering
    \includegraphics[width=1\linewidth]{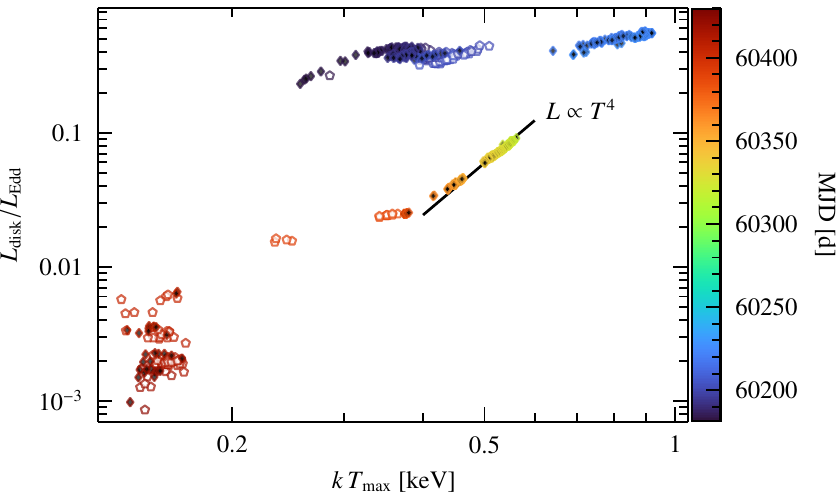}
    \caption{Disk luminosity as a function of temperature for all NICER data of the \swiftj outburst. The intrinsic disk flux is calculated from the \texttt{ezdiskbb} model (without Comptonization) in the 0.2--10\,keV range, and scaled to an Eddington luminosity using a distance of \distance and BH mass of $10\,M_\odot$. Diamonds and open pentagons denote orbit night and day data, respectively. The black line roughly going through the soft state data shows a Stefan–Boltzmann relation for a fixed radius.}
    \label{fig:kT_luminosity}
\end{figure}

In Fig.~\ref{fig:kT_luminosity}, we show the scaling of the disk luminosity as a function of temperature in analogy to studies using RXTE \citep[e.g.,][]{KubotaMakishima2004a,Done2007a,Gierlinski2008a,Dunn2011a} and Swift data \citep{ReynoldsMiller2013a} of other soft state LMXB BHs. 
We see that, during the soft state, the measured luminosity and temperature are strongly correlated and almost follow the Stefan-Boltzmann law (black line), but with a slightly shallower slope. Such deviations from the Stefan-Boltzmann law are often seen in the soft state, and are the result of temperature-dependent evolution of the color-correction. The effects of this dependency on the interpretation of disk truncation will be discussed in the next section. 
Finally, we note that in all other phases of the outburst of J1727, the temperature-dependent slopes deviate significantly from the Stefan-Boltzmann law, suggestive of other effects or changes in the system. 

\section{Discussion}
\label{sec:discussion}

NICER's coverage of the low-energy X-ray range combined with the high-cadence monitoring of J1727 allows us to track trend lines in disk luminosity, temperature, and radius with unprecedented precision (previous studies used instruments with limited sensitivity to the accretion disk emission, such as RXTE; see, e.g., \citealt{GierlinskiDone2004a,Gierlinski2008a,Dunn2011a}).
In this section, we show that with these high-quality data, taking temperature dependence in the color-correction treatment into account is strictly necessary. Otherwise, erroneous conclusions would be drawn regarding the disk's inner radius (consequently, on disk truncation). 
Specifically, we show that the soft state trends can be very well explained using the radiative transfer theory in standard thin disks \citep{DavisDoneBlaes2006a,DavisElAbd2019a}, yielding a constant radius associated with the ISCO. For the HIMS of J1727, we show that, with these model fits, variations in color-correction are not sufficient to interpret the disk at a constant radius \citep[e.g.,][]{Dunn2011a,Salvesen2013a}. However, we also discuss to what extent other systematics in the disk and Comptonization model can impact conclusions on truncation. In the back-transition, we find strong evidence for an onset of disk truncation right when the source leaves the soft state.

\subsection{Influence of a temperature-dependent color-correction factor on the inner disk radius: Soft state}
\label{sec:radius_temperature_plane}

The soft state spectra are dominated by the accretion disk that decreases in temperature from $kT_\mathrm{max}=0.56$ to 0.38\,keV (Fig.~\ref{fig:spectral_parameter_evolution}), while the luminosity in the 0.2--10\,keV band decays from roughly 4\% $L_\mathrm{Edd}$ to 1\% $L_\mathrm{Edd}$ (Fig.~\ref{fig:q_diagram}). The flux contribution from the corona is small ($f_\mathrm{sc}\lesssim 0.05$).
As motivated in the introduction, opacity effects in the disk atmosphere lead to a hardening of the disk radiation, a process that can be modeled by multiplying the effective temperature with $f_\mathrm{col}$  \citep{ShimuraTakahara1995a,DavisDoneBlaes2006a,DavisElAbd2019a}.

We reiterate that spectral hardening can have profound implications on the physical interpretation of the inner radius derived from disk continuum fitting.
For instance, \citet{MerloniFabianRoss2000a} showed that changes in the inner radius can be attributed to oversimplified disk models (e.g., \texttt{diskbb}) with constant $f_\mathrm{col}$. Instead, subtle trends in the disk radius can be readily explained by a change in $f_\mathrm{col}$ (see also, e.g., \citealt{KubotaMakishima2004a,GierlinskiDone2004a,Salvesen2013a,Sridhar2020a}).
From radiative transfer computations, the color-correction factor is found to be temperature-dependent and follows a proportionality of (see appendix of \citealt{DavisDoneBlaes2006a}\footnote{We note that there is a typo in the exponents of the $f_\mathrm{col}$ temperature relation in \citet{DavisDoneBlaes2006a}. The first line of page 537 says $f_\mathrm{col}\propto T_\mathrm{eff}^{1/4}\propto T_*^{1/3}$ but it should be $f_\mathrm{col}\propto T_\mathrm{eff}^{1/3}\propto T_*^{1/4}$. This typo becomes obvious from their Eq.~A5 (using their assumption of $\tau_* \sim T_*$).})
\begin{equation}
    \label{eq:f_col}
    f_\mathrm{col} \propto T_\mathrm{eff}^{1/3} \propto T_\mathrm{col}^{1/4} \quad . 
\end{equation}

\begin{figure}
    \centering
    \includegraphics[width=1\linewidth]{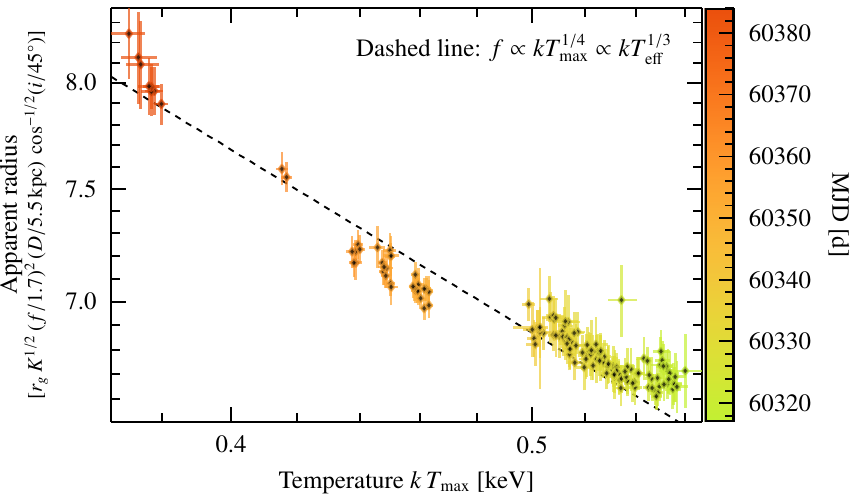}
    \caption{Apparent inner disk radius as a function of disk temperature for the soft state of \swiftj. The apparent radius is estimated from the disk normalization in the \texttt{simpl\textasteriskcentered ezdiskbb} model (Eq.~\ref{eq:norm_radius}), assuming a constant color-correction value of 1.7, a distance of \distance, and an inclination of $45^\circ$. A black hole mass of $10\,M_\odot$ is assumed for the conversion to $r_g$. The dashed line denotes a line of constant radius assuming the temperature-depending color-correction factor, $f_\mathrm{col}$, of \citet{DavisDoneBlaes2006a}. We associate this line of constant radius with the ISCO. Color denotes time with the same color-coding as in Fig.~\ref{fig:q_diagram} and \ref{fig:kT_luminosity}.}
    \label{fig:soft_state_fcol_fit}
\end{figure}

In the hard state, it is difficult to gauge the systematic uncertainties of the disk continuum fitting method due to the complex contribution from Comptonization.
Thus, we first evaluate the accuracy of the color-correction treatment of \citet{DavisDoneBlaes2006a} by testing the theoretical prediction against the data in the soft state of J1727, where the continuum fitting method is most reliable.
Figure~\ref{fig:soft_state_fcol_fit} shows the apparent inner disk radius of the soft state data as a function of disk temperature. These radii are ``apparent'' because they are calculated assuming a constant $f_\mathrm{col}=1.7$. In Fig.~\ref{fig:soft_state_fcol_fit}, a strong apparent evolution in radius is visible, contrary to theoretical and observational expectations of a constant disk at the ISCO \citep[e.g.,][]{TanakaLewin1995a,GierlinskiDone2004a,Steiner2010a}.
Next, we compare against the expected behavior when $f_\mathrm{col}$ is temperature-dependent according to Eq.~\ref{eq:f_col}, denoted as a dashed line in Fig.~\ref{fig:soft_state_fcol_fit}. Specifically, this corrected line of constant radius is calculated by
\begin{equation}
    \label{eq:f_proportionality}
    \frac{r_\mathrm{in,\,apparent}}{r_\mathrm{in,\,corrected}} = \left( \frac{f_\mathrm{const.}}{f_\mathrm{col}} \right)^2 \propto \left( \frac{1.7}{kT_\mathrm{max}^{1/4}} \right)^2 
\end{equation}
where we have used Eq.~\ref{eq:norm_radius} to take the ratio of the apparent radius obtained with a fiducial color-correction taken to be constant, $f_\mathrm{const.}$, to the corrected radius (normalization, distance, and inclination cancel out when taking the ratio). The proportionality arises from Eq.~\ref{eq:f_col} where we take the temperature-dependence of $f_\mathrm{col}$ into account.

It is remarkable how accurately the soft state data points follow the theoretically-predicted $f_\mathrm{col}$-proportionality of \citet{DavisDoneBlaes2006a} over a substantial range of disk temperatures\footnote{At high temperatures, the soft state data deviate slightly from the line, which may indicate a deviation from the standard disk to a slim disk.}. 
The data show clearly that a temperature-dependent $f_\mathrm{col}(T)$ must be taken into account for these high-quality soft state data. 
If one assumes a constant $f_\mathrm{col}$, one erroneously derives that there are significant changes in the inner radius. However, by accounting for $f_\mathrm{col}(T)$, we show that the soft state data of J1727 are entirely consistent with a constant radius, which we associate with the ISCO\footnote{In this treatment, we do not calculate exact values of $f_\mathrm{col}$ but only show the proportionality lines assuming Eq.~\ref{eq:f_col}. However, by anchoring the value of $f_\mathrm{col}(T)$ using a \texttt{kerrbb2} fit (see Sect.~\ref{sec:relativist_disk_models} for details), we can roughly estimate the range that it covers. We find that $f_\mathrm{col}(T)$ ranges from roughly 1.38 at $kT_\mathrm{max}=0.38\,\mathrm{keV}$ to 1.52 at 0.56\,keV.}. 

\subsection{Disk evolution in the HIMS}

\begin{figure}
    \centering
    \includegraphics[width=1\linewidth]{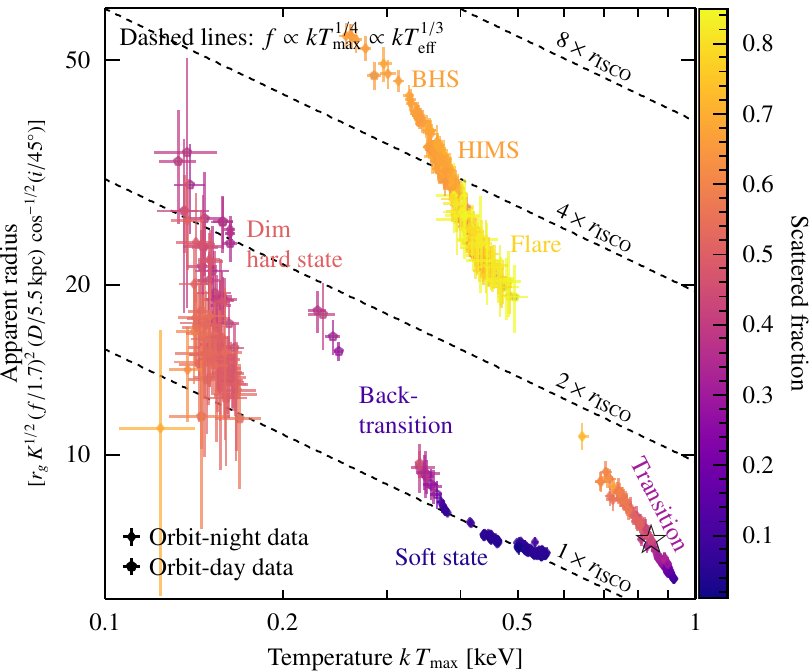}
    \caption{Apparent inner disk radius as a function of disk temperature for the full outburst of \swiftj. Data points are calculated as in Fig.~\ref{fig:soft_state_fcol_fit}. The color scale represents the amount of up-scattered photons in the corona, $f_\mathrm{sc}$, derived from \texttt{simpl}. Dashed lines denote lines of constant radius assuming a temperature-depending color-correction factor $f_\mathrm{col}$. Each line denotes an increase in radius by a factor of two. Black star in bottom right corner denotes the time of the hard-to-soft transition at MJD 60223.1 \citep{Hughes2025a}. A close-up of the soft state points is provided in Fig.~\ref{fig:soft_state_fcol_fit}.}
    \label{fig:ezdiskbb_rin_kt}
\end{figure}

We next apply the same analysis to the beginning of the NICER monitoring data, classified as the bright hard state, HIMS, and flare state (Sect.~\ref{sec:state_classification}). Figure~\ref{fig:ezdiskbb_rin_kt} shows the same radius-temperature plane as in the previous section, but incorporates data from all available states (effectively, Fig.~\ref{fig:soft_state_fcol_fit} is a zoom-in of Fig.~\ref{fig:ezdiskbb_rin_kt}). 
With this radius-temperature plane, it is possible to quantify the relative difference in inner radius without having to assume specific values on the distance, inclination, and BH mass (this only results in scaling on the y-axis), all whilst taking the temperature-varying $f_\mathrm{col}$ into account. We emphasize that, as values of the radius in gravitational radii depend on many parameters that are uncertain for \swiftj such as the mass and BH spin, we investigate trend lines and not absolute values of the disk radius. However, we make the implicit assumption that the $f_\mathrm{col}\propto T^{1/4}$ correlation holds for all states. We consider this potential source of systematics in Sect.~\ref{sec:model_systematics}.

The data from the first 25 days of the NICER monitoring follow a track from high apparent radius and low temperature to low apparent radius and higher temperature. To investigate whether the apparent radius trend is due to a temperature-dependent $f_\mathrm{col}$, we depict increasing lines of constant radius into this plane (using the proportionality in Eq.~\ref{eq:f_proportionality}) with each dashed line corresponding to a two-fold increase in radius. Thus, the second, third, and fourth lines from the bottom can be associated with $2\times r_\mathrm{ISCO}$, $4\times r_\mathrm{ISCO}$, and $8\times r_\mathrm{ISCO}$. 
The data exhibit a trend that is steeper than what can be explained by a constant radius.
Thus, the data imply the manifestation of inner-disk truncation well outside the ISCO, up to ${\lesssim}8\,r_\mathrm{ISCO}$. 

\subsection{Properties in the hard-to-soft transition}
\label{sec:hard-to-soft-transition}

The track of decreasing radius and increasing disk temperature throughout the HIMS continues into the data group labeled as hard-to-soft transition (bottom right of Fig.~\ref{fig:ezdiskbb_rin_kt}).
J1727 showed complex behavior during this period.  
Insight-HXMT and NICER data indicate that a Type-C quasi-periodic oscillation (QPO) is visible throughout the flare state \citep{Yu2024a}. \citet{Bollemeijer2023ATel16273} found a hard-to-soft transition to occur around MJD 60222 when the broadband noise changed and the QPO vanished (see also \citealt{Hughes2025a}). We depict this period by a black star in the bottom right corner of Fig.~\ref{fig:ezdiskbb_rin_kt}. Shortly after, on MJD 60223.1, a radio flare was detected 
\citep{MillerJones2023_ATel16271,Hughes2025a}. 

\begin{figure}
    \centering
    \includegraphics[width=1\linewidth]{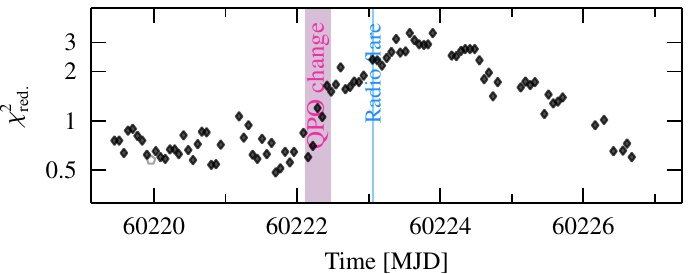}
    \caption{Evolution of the fit statistic at the time of the hard-to-soft state transition of \swiftj. The times of the QPO and broad band noise change from \citet{Bollemeijer2023ATel16273}, as well as the radio flare reported in \citet{Hughes2025a}, are indicated in magenta and blue, respectively.}
    \label{fig:transition_chi2}
\end{figure}

In our fits with \texttt{simpl\textasteriskcentered ezdiskbb}, we still achieve good fits up to around MJD 60222 (Fig.~\ref{fig:transition_chi2} and see example spectrum in Fig.~\ref{fig:spectra}). Beginning with the vanishing of the QPO, we find a significant spike in the fit statistic for about 4\,days. The coincidence of the change in the QPO over a few-hour span during which our otherwise successful spectral model performs very poorly\footnote{We note that we can achieve statistically acceptable fits when we let $N_\mathrm{H}$ vary by roughly 10\% (see Appendix Fig.~\ref{fig:absorption_estimate}). With such fits, other spectral parameters, however, follow trends that are counterintuitive, and we speculate that more complex physical processes occur during the state transition. A more detailed study of these spectra will be presented in a forthcoming paper.} suggests that the accretion flow deviates from a Comptonized standard thin disk (with fixed absorption) during this interval. 
As a result of the poor fits, caution is warranted and we do not rely on the model parameters for data between MJD 60222--60226 in arriving at our conclusions about J1727. Detailed fits exploring more complex spectral models for this interval are beyond the scope of this paper. 

\subsubsection{Onset of disk truncation in soft-to-hard transition}

\begin{figure}
    \centering
    \begin{tikzpicture}
        \node (p0) at (0,0){\includegraphics[width=1\linewidth]{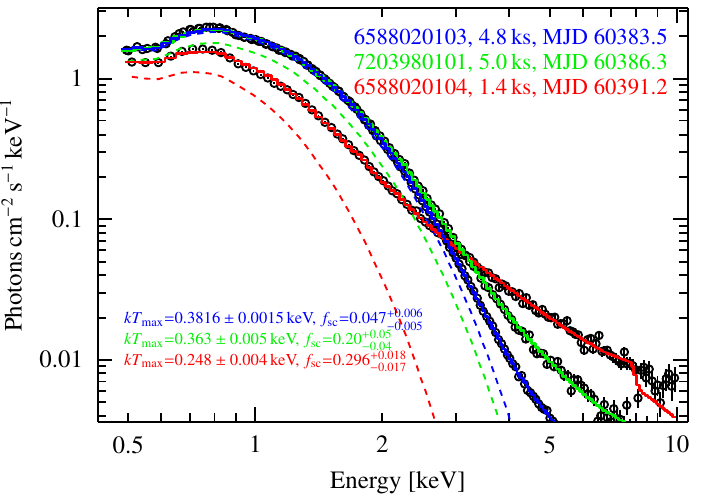}};
        \node (p1) at (2.6,.5){\includegraphics[width=.3\linewidth]{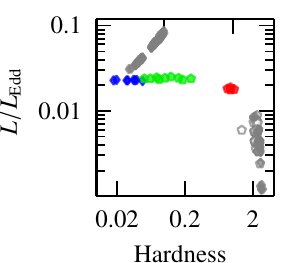}};
    \end{tikzpicture}
    \caption{Unfolded GTI-averaged spectra in the back-transition of \swiftj. Dashed lines show the contribution from the accretion disk ($\texttt{ezdiskbb}\cdot (1-f_\mathrm{sc})$, see Sect.~\ref{sec:luminosity}). Inset shows a zoom-in of the hardness-luminosity diagram in Fig.~\ref{fig:q_diagram} with the observations of the GTI-averaged spectra color-coded.}
    \label{fig:truncation_onset}
\end{figure}

After the three-month gaps in the NICER monitoring following the hard-to-soft transition, from MJD 60317.6--60384.0 (Obs. IDs 6203980138--6588020102), J1727 can be confidently associated with the soft state. In this section, we analyze the subsequent three Obs. IDs which occur during the transition from the soft state to the dim hard state. We show that we can derive a conclusion on disk truncation from the evolution in the hardness-luminosity diagram (Fig.~\ref{fig:q_diagram}), the radius-temperature plane (Fig.~\ref{fig:ezdiskbb_rin_kt}), and from the spectral shape using the GTI-averaged spectra of the three Obs. IDs (Fig.~\ref{fig:truncation_onset}). 

The blue spectrum in Fig.~\ref{fig:truncation_onset} shows the averaged spectrum of J1727 when it was just veering off of the faint end of the soft state (Obs. ID 6588020103, beginning on MJD 60383.5), exhibiting a temperature of 0.38\,keV, and a scattered fraction of roughly 0.05. This observation is still consistent with the $1\,r_\mathrm{ISCO}$ line (Fig.~\ref{fig:soft_state_fcol_fit}) and covers the left-most tip of the q-diagram, showing a sharp turn in hardness at MJD 60383.6. 

NICER's ability to monitor sources during ISS orbit-day allows us to constrain the evolution throughout the back-transition. Three days after the sharp turn in the q-diagram, beginning on MJD 60386.3 (Obs. ID 7203980101), significant differences in the disk and coronal properties can be seen with respect to the soft state. The scattered fraction, inferred from fitting the averaged spectrum, increases to around 0.20 while the disk cools to 0.363(5)\,keV. In the hardness-luminosity diagram, within the 14\,h of elapsed time in this observation, J1727 hardens by roughly one order of magnitude (from a hardness of 0.04 to 0.3), rapidly moving to the right in the q-diagram at constant luminosity. Furthermore, the tracks in the radius-temperature plane begin to deviate from the ISCO-line.

On MJD 60391.2 (Obs. ID 6588020104), the averaged spectrum shows the clear presence of a prominent Comptonized power law. In total, the source has hardened by roughly two orders of magnitude within the 7.6 days since the sharp turnaround in the soft state. The inferred radius from Fig.~\ref{fig:ezdiskbb_rin_kt} has roughly doubled. This behavior suggests the formation of a corona in the soft-to-hard transition, and the onset of mild disk truncation as the corona forms.

\subsection{Hysteresis behavior and comparison to \eighteentwenty}

The radius-temperature plane allows us to study the accretion flow in a ``physical source space'' rather than relying on properties such as the hardness-intensity diagram, which also contains detector properties. 
Due to the less extreme inner-disk truncation found in the dim hard state compared to the bright hard state, a hysteresis pattern with two parallel tracks is visible in Fig.~\ref{fig:ezdiskbb_rin_kt}. This behavior can be empirically connected to the hysteresis we see in the q-diagram of Fig.~\ref{fig:q_diagram}, with the upper track (HIMS through flare state through hard-to-soft transition) corresponding to the forward-transition, and the lower track corresponding to the backward-transition.

\begin{figure}
    \centering
    \includegraphics[width=1\linewidth]{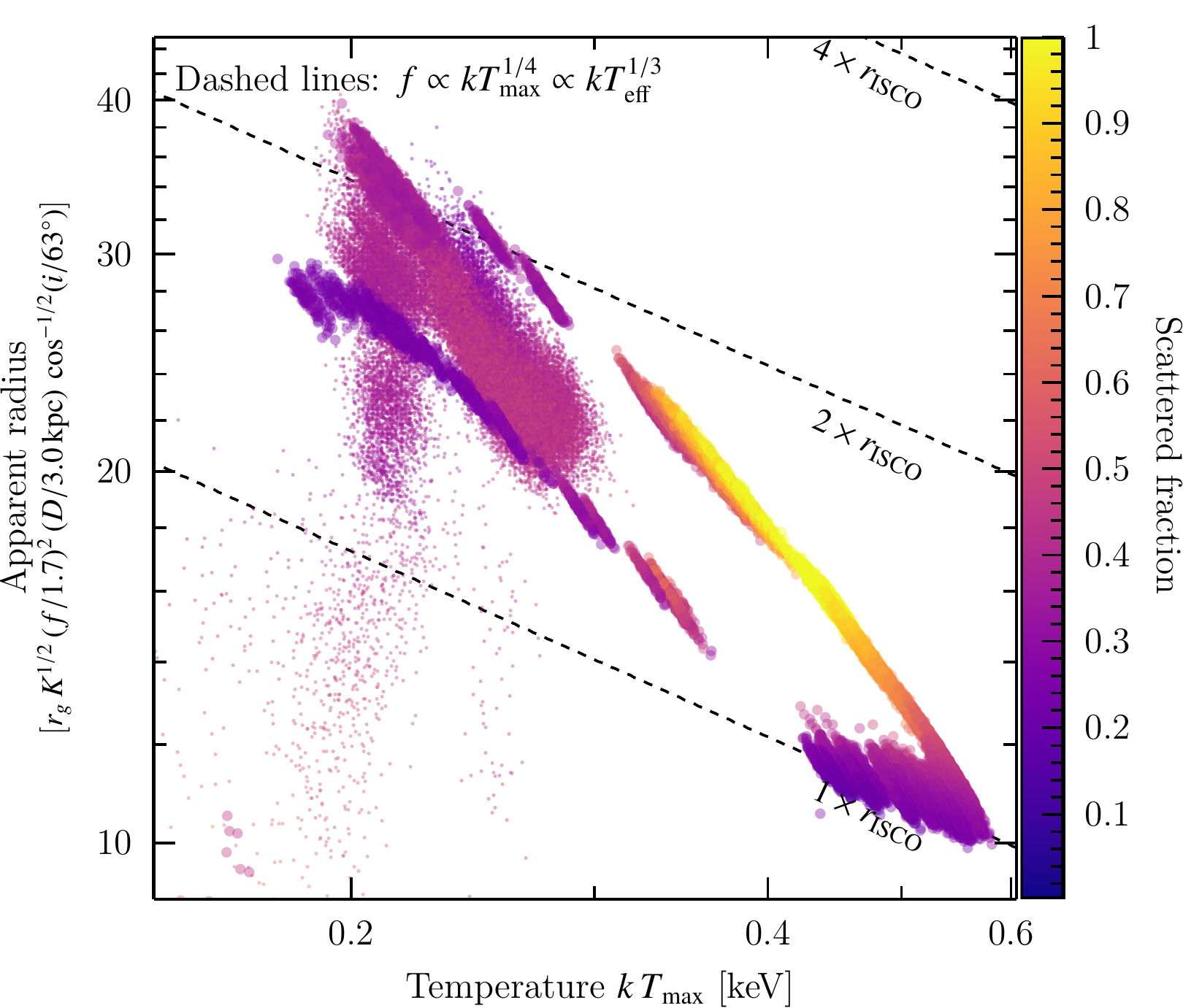}
    \caption{Apparent inner disk radius as a function of temperature for \eighteentwenty. Color coding and dashed lines are as in the corresponding plot for J1727 (Fig.~\ref{fig:ezdiskbb_rin_kt}). Note that the whole outburst evolution occurs between 1--$2\,r_\mathrm{ISCO}$, significantly closer to the black hole compared to J1727.}
    \label{fig:j1820_ezdiskbb_radius_kt}
\end{figure}

For the purposes of comparison with our results for J1727, we perform the same analysis on the 2018 outburst of J1820, and plot the resulting tracks of apparent inner radius versus disk temperature in Fig.~\ref{fig:j1820_ezdiskbb_radius_kt}. We identify a similar hysteresis pattern in the J1820 data. Again, two parallel tracks are visible, coinciding with the forward- and backward-transition.
However, the relative difference between the upper and lower tracks in J1820 is noticeably small compared to the difference from J1727 (also note that the relative difference in luminosity between the forward- and backward-transition is a factor of roughly 50 in J1727, while it is only a factor of roughly 5 in J1820, see Fig.~\ref{fig:q_diagram}). All J1820 data are within  $2\,r_\mathrm{ISCO}$. This more modest truncation amplitude makes an interpretation challenging as the model systematics (see Sect.~\ref{sec:model_systematics}) are sufficiently large that the apparent modest truncation in J1820 is in fact not very significant. This is because of large systematics for data outside of the soft/thermal state for which the scattered fractions is high and the uncertain geometry of the corona is most impactful on the radius inference. Nevertheless, it is interesting that J1820 shows a similar hysteresis trend to that of J1727. 

A hysteresis pattern similar to J1727 and J1820 has been found in \gx when modeled with the JED-SAD model \citep[][their Fig.~4]{Marcel2022a}. These authors attributed the hysteresis to changes in radiative efficiency. The main difference between their findings and our tracks in J1727 and J1820 is that we find a substantial difference in the apparent scale of truncation between the dim and bright hard states. \citet{Marcel2022a} find that the disk truncation is at the same level in the dim and bright hard states (see point E in their Fig.~4).

\subsection{Model systematics}
\label{sec:model_systematics}

The main challenge in the interpretation of our fits is understanding the underlying systematic uncertainties. The NICER data have a calibration uncertainty of ${\lesssim}2\%$, which belies the larger possible impact of systematic modeling uncertainties. In fact, the data quality allows us to identify very clear trends across the whole dataset, with extremely low scatter, as seen in Fig.~\ref{fig:ezdiskbb_rin_kt}. Quantifying the systematics from the model is more challenging. 

\subsubsection{Systematics arising from uncertain coronal geometry}

For the corona, the main source of uncertainty is not the process of Comptonization itself, believed to be described well in the models \citep{Zdziarski1996a,Zycki1999a,Steiner2009a,Zdziarski2020a}, but the assumption of the coronal geometrical configuration \citep[see also Fig.~13 in][]{Done2007a}. 
Depending on the fraction of the disk covered by the corona, or the structure of the corona (e.g., clumpiness or density variance), a spectral model's inference of the underlying seed spectrum will change in amplitude and, thus, the inferred radius changes (according to Eq.~\ref{eq:norm_radius}).
In order to conservatively estimate the contribution on the error budget, we use the scattered fraction, $f_\mathrm{sc}$, of the \texttt{simpl} model, denoting the fraction of seed photons being Comptonized and adopt a conservative order unity uncertainty on the Compton-scattering normalization arising from uncertainty in coronal geometry. The corresponding fractional systematic uncertainty on radius is $(f_\mathrm{sc}/(1-f_\mathrm{sc}))^{1/2}$. This uncertainty is meager, ${<}5\%$, in the soft state, but up to 150\% in the HIMS, where $f_\mathrm{sc}\sim 0.7$ (see Fig.~\ref{fig:spectral_parameter_evolution}). 
In the back-transition, where $f_\mathrm{sc}\sim 0.3$, we estimate a systematic uncertainty on the radius of around 65\%, respectively. Figure~\ref{fig:ezdiskbb_rin_kt} suggests truncation by a factor of 4 in the HIMS, and a factor of 2 in the back-transition, both of which significantly exceed our highly conservative error estimate. Thus, we consider the appearance of inner-disk truncation to be significant.

\subsubsection{Systematics arising from relativistic accretion disk models}
\label{sec:relativist_disk_models}

The error budget is also influenced by the choice of the accretion disk model. The model employed in Sect.~\ref{sec:results}, \texttt{ezdiskbb}, neglects relativistic effects. Specifically, Doppler broadening leads to a broader disk continuum compared to the non-relativistic treatment, particularly for highly inclined sources \citep{Straub2011a}. We test the influence of the relativistic correction by fitting the models \texttt{kerrbb} \citep{Li2005a}, \texttt{kerrbb2} \citep{McClintock2006a}, which both have a fixed inner radius, and \texttt{kynbb} \citep{Dovciak2004a}, which has the radius as a free parameter. Each disk model is Comptonized with \texttt{simpl}.

\begin{figure}
    \centering
    \begin{tikzpicture}
    \node (p0) at (0,0){\includegraphics[width=1\linewidth]{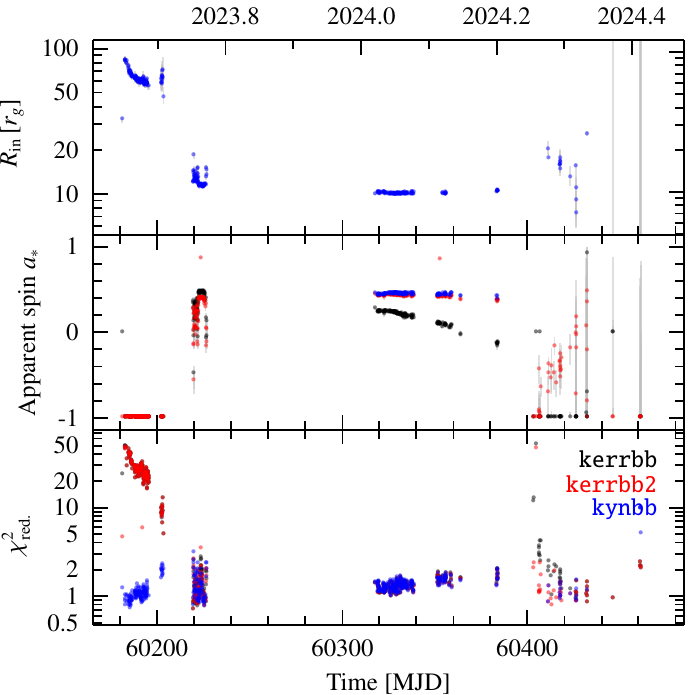}};
    \node (p1) at (-.3,-.3){\includegraphics[width=.25\linewidth]{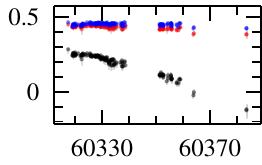}};
    \draw[thick,gray] (-.95, .31) -- (.3, .7);
    \draw[thick,gray] (.75, -.65) -- (1.8, -.05);
    \end{tikzpicture}
    \caption{Parameter evolution of \swiftj throughout its 2023/2024 outburst fitted with the relativistic disk models \texttt{kerrbb} (black), \texttt{kerrbb2} (red), and \texttt{kynbb} (blue). Only orbit-night data points are shown. In the top panel, the \texttt{kynbb} models assumes a fixed black hole spin of 0.9 (see text for further explanation of this choice). The middle panel assumes $R_\mathrm{in}=R_\mathrm{ISCO}$ for all models. In the inset, we show soft state fits with a radius fixed to the ISCO and free spin. We assume a fiducial black hole mass of $10\,M_\odot$, an inclination of $45^\circ$, and a distance of \distance. In all fits, the Comptonization model is \texttt{simpl}, and the absorption is fixed to $N_\mathrm{H} = 2.84\times 10^{21}\,\mathrm{cm}^{-2}$.} \label{fig:spectral_parameter_evolution_relativistic_disk_models}
\end{figure}

The middle panel of Fig.~\ref{fig:spectral_parameter_evolution_relativistic_disk_models} shows the spin evolution of the relativistic disk models (the coronal parameters, $\Gamma$ and $f_\mathrm{sc}$, are consistent with Fig.~\ref{fig:spectral_parameter_evolution}). 
The \texttt{kerrbb/2} models are designed to infer spin from an inner radius at the ISCO that is not free to vary (at a fixed unity normalization). Instead, one can allow the spin to vary as a proxy for a change in truncation. A lower spin means larger truncation because a non- or negatively-spinning black hole has a larger ISCO radius than a maximally-rotating black hole. This technique works for truncation radii as large as $9\,r_g$ ($a_*=-1$; however, see also \citealt{Mummery2024b}). If the disk is truncated further outside, the \texttt{kerrbb/2} disk models no longer accommodate this radius range and so result in poor-quality fits to the data. 
Consequently, the sizeable truncation that we inferred in the HIMS in Sect.~\ref{sec:radius_temperature_plane} can be identified in Fig.~\ref{fig:spectral_parameter_evolution_relativistic_disk_models} as a trend reaching large $\chi_\mathrm{red.}^2$ values at the beginning of the monitoring data (before MJD 60200).

Differences between \texttt{kerrbb} and \texttt{kerrbb2} become most visible in the soft state where the disk can be assumed to reach the ISCO. 
The \texttt{kerrbb} model does not self-consistently calculate the color-correction factor, and any temperature dependence has to be supplied manually. If $f_\mathrm{col}$ is kept at a constant value, the spin shows a clear drift by ${\sim}25\%$ throughout the decaying soft state (see inset in Fig.~\ref{fig:spectral_parameter_evolution_relativistic_disk_models}). 
As explained in Sect.~\ref{sec:radius_temperature_plane}, the $f_\mathrm{col}\propto T^{1/4}$ dependence needs to be taken into account to obtain a meaningful and (empirically) constant radius. For the \texttt{kerrbb2} fits, this dependence is accounted for internally, and the spin values for the soft state are remarkably flat at 0.45. The scatter is only around 1--2\% across the three-month-long soft state observations. (We note that we do not aim to provide a proper spin estimate due to a lack of precise knowledge of the  black hole's mass, inclination, and distance; rather, we highlight the stability of the soft state fitting here; the constrained value is smaller than $a\approx 0.9$, reported by \citealt{Svoboda2024a}, which is due to the larger distance assumed in this analysis -- if we assume a smaller distance of 3.4\,kpc, we obtain $a\approx 0.87$). 

\begin{figure}
    \centering
    \includegraphics[width=1\linewidth]{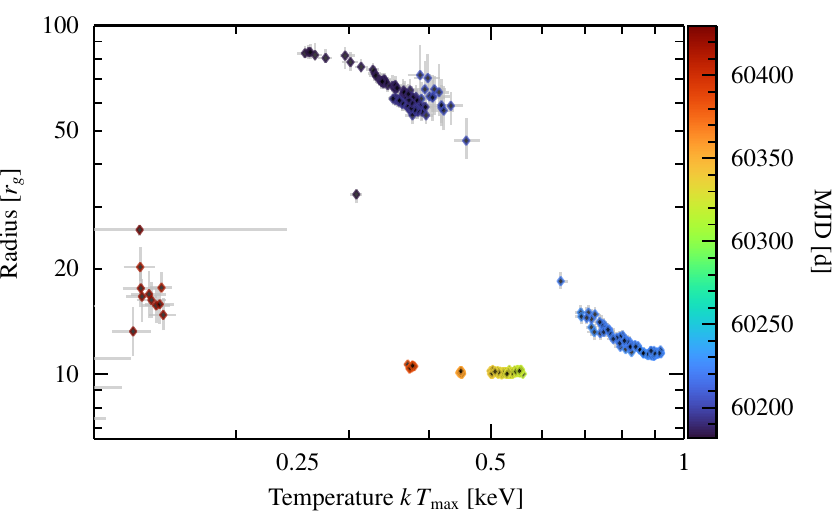}
    \caption{Inner disk radius from the relativistic disk model \texttt{kynbb} as a function of temperature for the full 2022/23 outburst of \swiftj. The plot is analogous to Fig.~\ref{fig:ezdiskbb_rin_kt} but includes relativistic effects in the radius estimation. Only orbit-night data points are shown. The temperature is inferred from the \texttt{ezdiskbb} fits. In the \texttt{kynbb} fits, only the mass accretion rate and the inner radius are fitted. The spin is fixed to 0.9, and a black hole mass, inclination, and distance of $10\,M_\odot$, $45^\circ$, and \distance are assumed. The color-correction factor fixed to reflect the $f_\mathrm{col}\propto T^{1/4}$ scaling, where we obtain the temperature from the corresponding \texttt{ezdiskbb} fit. A hysteresis behavior is visible, as for the non-relativistic model fits.}
    \label{fig:kynbb_rin_kt}
\end{figure}

\texttt{kynbb} does not provide the $f_\mathrm{col}\propto T^{1/4}$ scaling, so we have to manually set $f_\mathrm{col}$. In the soft state, we infer a color-correction factor of $f_\mathrm{col}=1.577$ from \texttt{kerrbb2} for a disk temperature of $T_\mathrm{max}=0.55$\,keV (inferred from Obs. ID 6203980142). For the \texttt{kynbb} fits, we use this value as an anchor point, and then manually scale $f_\mathrm{col}=1.577\cdot T_\mathrm{max}^{0.25}/0.55^{0.25}$, using the corresponding temperature from the \texttt{ezdiskbb} fit.
After applying this correction and fixing the radius to the ISCO in the soft state, the spin evolution of \texttt{kynbb} is flat and consistent with \texttt{kerrbb2} (see again inset of Fig.~\ref{fig:spectral_parameter_evolution_relativistic_disk_models}).

To model the full outburst with \texttt{kynbb}, we fix the spin and let the radius free (upper panel of Fig.~\ref{fig:spectral_parameter_evolution_relativistic_disk_models}). A spin value of 0.9 was adopted to avoid a fit-convergence problem in \texttt{kynbb}. At the best-fit spin of a=0.45, allowing the inner radius, $R_\mathrm{in}$, to vary freely, does not let the model distinguish between $R_\mathrm{in}=R_\mathrm{ISCO}$ and $R_\mathrm{in}<R_\mathrm{ISCO}$. This is because \texttt{kynbb} does not assume any additional emission inside the ISCO, as the zero-torque boundary condition is fixed there. As a result, fits with free $R_\mathrm{in}$ converge poorly at this spin. Choosing a substantially higher spin resolves this issue by making $R_\mathrm{ISCO} < R_\mathrm{in}$. This allows the radius to be constrained more precisely and makes it possible to test its stability over time and across changes in luminosity. We will further elaborate on this behavior in Sect.~\ref{sec:zero-torque_inner_boundary_condition} when we discuss the impact of the inner disk boundary condition.

Using these inner radius measurements, we then plot the radius-temperature plane, analogous to the \texttt{ezdiskbb} treatment, in order to investigate whether the truncation evolution and hysteresis persists in the relativistic treatment. 
Figure~\ref{fig:kynbb_rin_kt} shows that the disk is significantly truncated in the HIMS, reaches the ISCO at around the time of the state transition, is then roughly constant in the soft state, and becomes truncated as J1727 exits the soft state. This interpretation is qualitatively consistent with what we inferred from the non-relativistic case. However, we note that there is a difference in the absolute values of the truncation. In the HIMS, \texttt{kynbb} suggests around 11--18\,$r_\mathrm{ISCO}$ (50--80\,$r_g$; roughly consistent with \citealt{Ma2025a}), while \texttt{ezdiskbb} suggests around $6\,r_\mathrm{ISCO}\approx 30\,r_g$ for $a\approx 0.45$. This factor of ${\approx}2$ difference may be due to the aforementioned relativistic Doppler broadening of the disk emission, or the inner disk boundary condition. 

It is reasonable to consider that the J1727 data during the HIMS consists largely of emission from the radiation pressure-dominated regime due to its high luminosity. Therefore, we also test the parameter evolution with the \texttt{slimbh} model \citep{Sadowski_PhD_2011}, which also assumes an inner disk radius at the ISCO. We find consistent behavior to the \texttt{kerrbb/2} fits, i.e., the spin pegs at the minimally allowed value, consistent with truncation. We do not include these results because the \texttt{slimbh} can only be applied at ${>}5\%\,L_\mathrm{Edd}$, which does not cover most of the soft state and the back-transition. 

In summary, we find that most currently used relativistic disk models have limitations in their dynamical parameter ranges and are unable to model the full dataset of J1727. This can be due to the inability to decouple the inner radius from the ISCO (for \texttt{kerrbb/2} or \texttt{slimbh}, radius changes are modeled by proxy as spin changes, which can only vary in a limited range) or the dynamical range of the luminosity (for \texttt{slimbh}, the luminosity must be greater than $5\% L_\mathrm{Edd}$, and in \texttt{kerrbb2}, $f_\mathrm{col}$ is only tabulated in a limited luminosity range). In \texttt{kynbb}, the radius is a free parameter, which makes it the most versatile model for our analysis. However, it requires a manual setting of $f_\mathrm{col}\propto T^{1/4}$ and its parameterization of the inner disk boundary condition appears to lead to complex convergence behavior. The complexity of the J1727 outburst motivates further development of these disk models.

\subsubsection{Systematics arising from the color-correction parametrization}
\label{subsec:color-correction}
Much of the interpretation on disk truncation from Fig.~\ref{fig:ezdiskbb_rin_kt} relies on the assumption that the $f_\mathrm{col}$ temperature correlation is applicable in the hard state. This is an important assumption we make. It is as of yet unverified by theory, but we note that, e.g., accounting for vertical structure in the disk or considering an energetically dominant corona in the inner disk vicinity (e.g., in a slab geometry) could conceivably affect the color correction. Indeed, some studies have proposed an evolution in color-correction to be responsible for the apparent disk truncation \citep[e.g.,][]{Salvesen2013a,SalvesenMiller2021a}. With the NICER data of J1727, we can quantitatively evaluate what this would imply for the dependence of color and effective temperature from a data-driven perspective. 

Both of the transition tracks in Fig.~\ref{fig:ezdiskbb_rin_kt} can be reasonably described by power-laws $r_\mathrm{apparent} \propto T_\mathrm{max}^\gamma$.  We fit directly to the data and infer $\gamma_\mathrm{forward}=-1.870\pm 0.011$ for the forward transition to the soft state and a slightly shallower backwards transition to the dim hard state $\gamma_\mathrm{backward}=-1.60\pm 0.15$ (for details on the fits, see Appendix Fig.~\ref{fig:app:fit_tracks}). Using Eq.~\ref{eq:f_proportionality}, which shows $r_\mathrm{apparent} \propto f_\mathrm{col}^{-2}$, we now explore the possibility that the complete track can be explained by a changing color-correction at a constant radius. 
The $f_\mathrm{col}(T)$ dependence then obeys $f_\mathrm{col} \propto T_\mathrm{col}^{-\gamma/2}$. Combining this dependence with the definition of color-correction ($f_\mathrm{col}=T_\mathrm{col}/T_\mathrm{eff}$) and solving for $T_\mathrm{col}$, we infer $T_\mathrm{col} \propto T_\mathrm{eff}^{1/(1+\gamma/2)} \propto L^{1/(4+2\gamma)}$. For our fitted $\gamma \approx -1.5$ to $-2$, we conclude that $T_\mathrm{col}$ would have to increase extremely sensitively with $T_\mathrm{eff}$ (or analogously, with luminosity). In other words, the color temperature must shift significantly at a nearly constant luminosity. Any disk atmosphere model that maintains constant radius during the transitions would need to fulfill this requirement (at least for J1727), and to our knowledge no model predicts such behavior.

\citet{DavisDoneBlaes2006a} outline that there are two regimes, photon-saturated ($f_\mathrm{col}\propto T^{1/4}$) at the low temperature end, and photon-starved ($f_\mathrm{col} \propto T^{-1/9}$) at very high temperatures. These calculations do not take radiation from the corona into account. From the fitted $\gamma_\mathrm{forward}$, in principle, we can indeed align the data points of the complete forward-transition on a line of constant radius for an approximately $f_\mathrm{col} \propto T^1$ scaling. Note that in such a regime ($\gamma \sim -2$) requires extremely steep dependency of $T_\mathrm{col}$ with $L$ (roughly as $L^4$ for the forward transition). This would fundamentally change the interpretation of this paper.

One argument against such a significant modification of the $f_\mathrm{col}(T)$ dependence is that if there is a significant change of this dependence in the hard state relative to the soft state, we would expect the slope of the radius-temperature tracks to correlate with $f_\mathrm{sc}$, which parameterizes the degree of Comptonization. However, Fig.~\ref{fig:ezdiskbb_rin_kt} and \ref{fig:j1820_ezdiskbb_radius_kt} show two essentially parallel tracks, with no clear bend except for the sharp turning point coinciding with the hard-to-soft transition, and the onset of truncation as the source exits the soft state\footnote{As J1727 and J1820 decay into quiescence, Figs.~\ref{fig:ezdiskbb_rin_kt} and \ref{fig:j1820_ezdiskbb_radius_kt} show that some data points have a trend towards lower radii. This behavior is inconsistent with \citet{RemillardMcClintock2006a} where the accretion disk is expected to be highly truncated in quiescence. The trend may arise from inaccuracies in modeling due to appreciable background contamination at high energies when at low fluxes, or due to a more complex continuum towards quiescence.}. We suggest that a detailed exploration of this behavior from a theoretical perspective is imperative in order to reach a firm conclusion on its physical relevance here, or lack thereof. Such an investigation is outside the scope of this work. 

Furthermore, \citet{Fukumura2025a} have recently argued that in the presence of strong disk winds, i.e., particularly in the soft state, Compton scattering of disk photons in the wind and, thus, down-scattering of radiation, presents an effect counteracting the disk atmosphere scattering treated by $f_\mathrm{col}$. They conclude that the color-correction treatment may not be sufficient in this case. We do not see strong absorption lines in the J1727 soft state spectra that would point to a disk wind, which may be explained by J1727 likely not being highly inclined (although we cannot exclude that an ionized wind is present). In addition, we show that the soft state data follow the theoretically expected color-correction very accurately. Disk winds would generally be expected to have a luminosity dependence, and so the fact that the whole soft state track appears unaffected (i.e., flat in radius) weighs appreciably against such effects mattering in J1727.

\subsubsection{Systematics from the inner disk boundary condition}
\label{sec:zero-torque_inner_boundary_condition}
In Sect.~\ref{sec:radius_temperature_plane}, we have used the \texttt{ezdiskbb} model, which has a zero-torque boundary condition. In the presence of truncation, e.g., if a hot inner flow is present in the HIMS, angular momentum is transported through the inner disk radius and this assumption is incorrect. 
As we tentatively find significant truncation in the HIMS, we deploy the \texttt{diskbb} model, which has no zero-torque boundary condition, to see if this affects the overall conclusions of this paper. In Appendix Fig.~\ref{fig:app:diskbb_rin_kt} one can see that the relative truncation levels remain similar, and our main finding (large-scale truncation evolution throughout the outburst) is not influenced by this choice.
When considering the apparent radius values (that is, no $f_\mathrm{col}(T)$ correction), however, we caution that the inner disk boundary does significantly change the absolute values by a factor of around 2 (compare y-axis values of Figs.~\ref{fig:ezdiskbb_rin_kt} and \ref{fig:app:diskbb_rin_kt}).  

Particularly in the soft state, the torque boundary condition is related to the question of whether there is material within the ISCO that can radiate energy efficiently.
Such emission from the ``plunging region'' has been speculated to exist (e.g., \citealt{Penna2010a,Zhu2012a,WilkinsReynoldsFabian2020a,Fabian2020a,Wilkins2021a,Lasota_2024arXiv,Mummery2025a}; Demiroglu et al. in prep.). 
We find no evidence of a strong quasi-thermal component in the 6--10\,keV region in J1727, as reported in \citet{Fabian2020a} in the source J1820 \citep[see also][]{Mummery2024a}. However, we found a peculiar effect in the \texttt{kynbb} fits of the J1727 soft state data. Firstly, we could not achieve stable fits for a=0.45, which was suggested by \texttt{kerrbb2}. Furthermore, the fitted inner disk radius values for a=0.9, shown in Fig.~\ref{fig:spectral_parameter_evolution_relativistic_disk_models} (upper panel), are not at the ISCO for an a=0.45 black hole (${\sim}4.4 r_g$), but are larger, around $10 r_g$. We suspect this is a result of the zero-torque boundary condition being applied at the ISCO for the adopted spin (about $2.5\,r_g$ for a=0.9) by  \texttt{kynbb}. For larger radii, the emissivity profile becomes discontinuous, vanishing abruptly at the inner-edge. This likely accounts for the factor ${\gtrsim}2$ difference between the best-fitting inner-radius here and the ISCO associated with the best-fitting spin from \texttt{kynbb} fits which assume the inner-radius reaches the ISCO. This complexity indicates the importance of a full exploration of accretion disk models from a data-driven perspective. Such efforts are ongoing, and further exploration of \texttt{kynbb} as well as applications of models like \texttt{fullkerr} \citep{Mummery2024a} will be valuable in studying the effects of the boundary conditions on radius determination for different states and regimes.

\subsubsection{Systematics arising from relativistic reflection and complex coronal continua}
\label{sec:relativistic_reflection}
Our analysis uses an empirical approach to model the iron line region with the \texttt{laor} model, aiming to achieve flat residuals in the NICER passband. Broadband analysis of J1727, including hard X-rays, suggests complex, potentially multi-component plasma regions and/or reflection (\citealt{2024Liu_arXiv}; Mastroserio et al., in prep.). While it is technically more correct to use a modern reflection model (see, e.g., Fig.~2 of \citealt{Dauser2010a}), there is currently no model available to self-consistently incorporate the varying accretion disk temperature across the outburst (in \texttt{relxill}, the seed temperature for the underlying \texttt{nthcomp} illumination used to calculate the reflection features is fixed to 50\,eV), which influences the low-energy passband (Fig.~3 in \citealt{Ubach2024a}). A solution is to artificially suppress the low-energy reflection spectrum by multiplying \texttt{relxill} with a broken power-law (Sect.~3.2 in \citealt{Ubach2024a}). However, as we find that reflection is not a dominant contribution to the J1727 spectrum in the NICER passband, we decided to avoid distraction from the main aspect of this paper (the behavior of the accretion disk) by over-complicating the reflection model. This may, however, introduce slight biases in the disk continuum parameters since we do not consider reflection at soft X-rays.

Similarly, biases in our results can arise if multiple coronal regions with different plasma properties, and thus different continua, are present \citep[e.g.,][and references therein]{Zdziarski2021a}. However, more complex coronal models are difficult to constrain with NICER alone due to the lack of hard X-ray coverage, and we do not explore the effects of different coronal modeling on the disk parameters.

\section{Summary and conclusions}
\label{sec:conclusions}

In this paper, we investigate changes of the accretion disk inner edge in \swiftj using disk continuum fitting of NICER data. We interpret relative changes in radius rather than focusing on absolute values, and trace the disk evolution throughout the outburst in a physical radius-temperature plane.

In the forward-transition of J1727 (bright hard state and HIMS), the data are consistent with substantial truncation, exhibiting a drift towards decreasing radii. In the soft state, we find the disk radius to be stable, consistent with the interpretation that the disk edge is at the ISCO.
When the source leaves the soft state, the spectral shape changes significantly, visible as a sharp pivot in hardness. This point coincides with the formation of a corona and the onset of disk truncation, which can be seen as a slight but significant increase in radius as the source back-transitions into the dim hard state. The back-transition covers roughly eight days.
Furthermore, the level of truncation in the back-transition is smaller than in the forward-transition, creating a hysteresis pattern in the radius-temperature plane.
We show that a similar hysteresis, albeit on a smaller scale, can be found in the LMXB black hole \eighteentwenty. 

Large parts of this paper are dedicated to assessing sources of systematic uncertainties that are important to treat in the disk models and can potentially lead to wrong inferences if not accounted for.
We find that the color-correction treatment has a significant influence on the interpreted results. This is best illustrated in the soft state where an apparent drift in radius occurs that vanishes if we account for the theoretically predicted temperature dependence ($f_\mathrm{col}\propto T^{1/4}$; \citealt{DavisDoneBlaes2006a}).
We also assess most currently available relativistic disk models and find that no model satisfactorily describes the full outburst of J1727, either because the models run out of dynamical range in their radius parameterization (\texttt{kerrbb/2}, \texttt{slimbh}) or because of its inner disk boundary condition (\texttt{kynbb}). 
To the extent possible in comparing the results across these disk models, the truncation trends and hysteresis behavior are consistent.

This study shows that the disk continuum fitting technique can be pushed into the hard states, although assessing the systematics becomes increasingly difficult due to an uncertain coronal geometry. Interpreting relative changes of the disk throughout time helps to understand these systematics at an unprecedented level, largely owing to the low-energy coverage and high-cadence monitoring capability of the NICER telescope.
Understanding the underlying physical mechanism of the different truncation levels in the dim and hard states, however, requires dedicated future theoretical modeling. 

\begin{acknowledgments}
We thank the anonymous referee for useful comments that improved the discussion on the impact of color-correction on disk truncation.
OK thanks James-Miller Jones for valuable discussions on the distance estimates of this source.
OK acknowledges NICER GO funding 80NSSC23K1660.
This research has made use of a collection of ISIS functions (ISISscripts) provided by ECAP/Remeis observatory and MIT (\url{http://www.sternwarte.uni-erlangen.de/isis/}). 
TD acknowledges support from the Deutsche Forschungsgemeinschaft (DFG, German Research Foundation) as part of the DFG Research Unit FOR5195 – project number 443220636.
MD and JS acknowledge support from the grant 26-22614S.
YZ acknowledges support from the Dutch Research Council (NWO) Rubicon Fellowship, file no.\ 019.231EN.021. AI is funded by the European Union (ERC, X-MAPS, 101169908). Views and opinions expressed are however those of the author(s) only and do not necessarily reflect those of the European Union or the European Research Council. Neither the European Union nor the granting authority can be held responsible for them.
\end{acknowledgments}

\vspace{5mm}
\facilities{NICER}

\software{Slang/ISIS \citep{Houck2002a},  
          HEASOFT (NICERDAS)
          }

\bibliography{references}{}
\bibliographystyle{aasjournal}

\appendix
\renewcommand\thefigure{\thesection.\arabic{figure}} 

\section{NICER noise peak fitting: Low-energy behavior of \swiftj and other black hole transients}
\label{sec:app:noisepeak}
\setcounter{figure}{0}

\begin{figure}[H]
    \centering
    \includegraphics[width=0.32\linewidth]{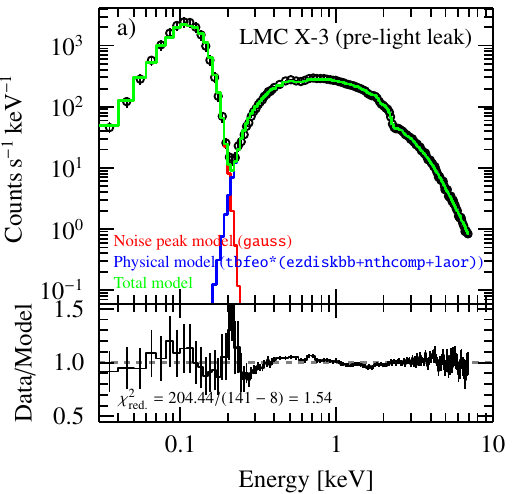}
    \hfill
    \includegraphics[width=0.32\linewidth]{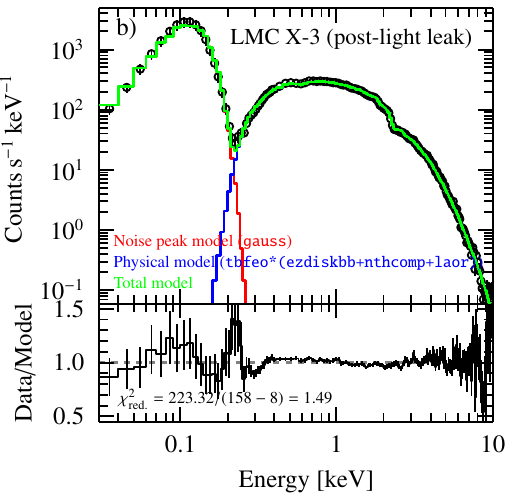}
    \hfill
    \includegraphics[width=0.32\linewidth]{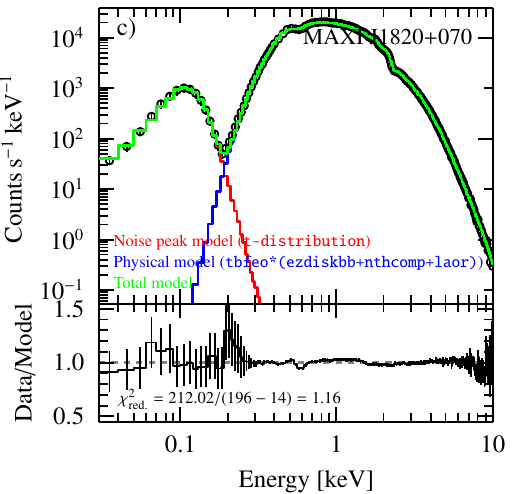}
    \caption{NICER spectral fits of soft state black hole X-ray binaries with our noise peak model. \textit{Left:} LMC~X-3 soft state observation 2101010104 from before the light leak, with the noise peak modeled with a Gaussian distribution. \textit{Center:} LMC~X-3 soft state observation 6101010108 from after the light leak event. \textit{Right:} \eighteentwenty soft state state observation 1200120203.}
    \label{fig:BHXRB_noisepeak_study}
\end{figure}

\begin{figure}
    \centering
    \includegraphics[width=0.5\linewidth]{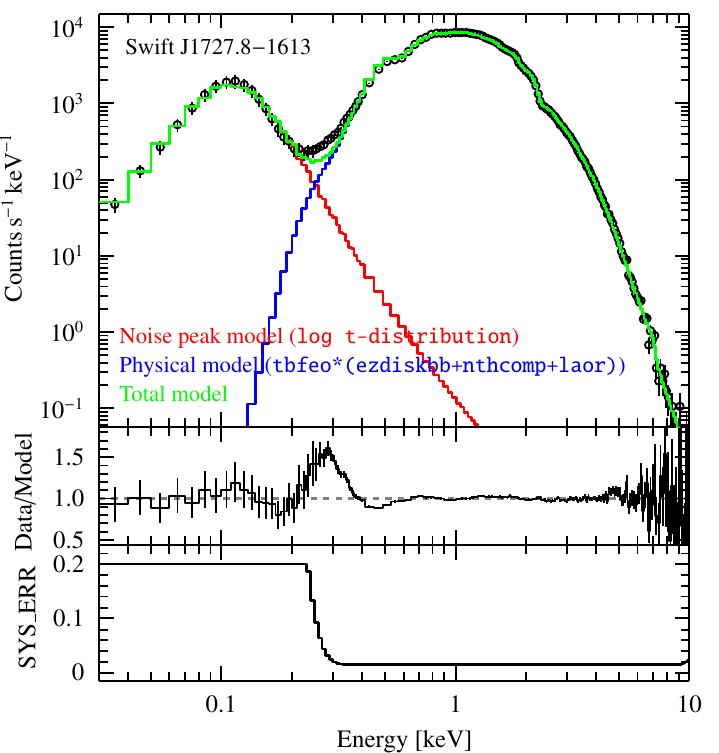}
    \caption{Noise peak modeling of \swiftj soft state observation 6203980141. Large residuals can be seen below roughly 0.45\,keV that cannot be described with our noise peak model.
    The bottom panel shows the systematic uncertainty that we apply to the data.}
    \label{fig:js_ni6203980141_0mpu7_dark_GTI1_noisepeak_fit}
\end{figure}

NICER's detectors each produce a roughly Gaussian-shaped noise peak at low energies that originates from the detector resets (undershoots). The noise peak is usually centered around 0.12\,keV and has a width of ${\sim}35$\,eV for moderate undershoot rates\footnote{\url{https://heasarc.gsfc.nasa.gov/docs/nicer/analysis_threads/noise-ringers/}}. If the undershoot rate is large, e.g., due to optical loading, the noise peak increases in amplitude and width, and can intrude higher energies. 
In this section, we fit the noise peak and soft state spectrum of the sources \eighteentwenty, LMC~X-3 (Fig.~\ref{fig:BHXRB_noisepeak_study}), and \swiftj (Fig.~\ref{fig:js_ni6203980141_0mpu7_dark_GTI1_noisepeak_fit}) with the aim of determining a common lower energy bound in our J1727 fits.

The noise peak is a detector property and does not arise from photons going through the mirrors. To model it, we use a flat effective area that is set to unity. As the noise peak is also not part of the detection process in the silicon drift detectors, we use a diagonal redistribution matrix with a resolution of 1\,eV such that we can sample the low-energy range properly.
With this approach, we can divide our model into two additive parts, a detector model, that uses the diagonal ``dummy'' response and a physical model, which uses the usual data response. 
Our detector model is either a Gaussian or a slightly broader t-distribution (the latter is motivated by the possibility that few individual detectors may have a substantially broader noise peak, which would broadened the total spectrum).
Our physical model is \texttt{tbfeo*(ezdiskbb+nthcomp*laor)}.
We do not rebin the noise peak range. Above 0.45\,keV, we use the optimal binning scheme of \citet{KaastraBleeker2016a} and fit the combined data using a $\chi^2$-statistic.
Further, we do not include background in the noise peak fits, as the SCORPEON model accounts for a Gaussian-shaped noise peak, that we here model ourselves.
To account for systematic uncertainties, we use the \texttt{SYS\_ERR} column that is written into the spectral file during \texttt{nicerl3-spect}. The column has a value of 1.5\% across most energies and increases around 0.2\,keV. We limit the maximum systematic uncertainty at 20\% in the noise peak range (see Fig.~\ref{fig:js_ni6203980141_0mpu7_dark_GTI1_noisepeak_fit}, \textit{lower panel}).

For the LMC~X-3 and \eighteentwenty soft state observations, we observe that the 0.03--10\,keV range can be described well with a Gaussian or slightly broadened t-distribution representing the noise peak. 
We test whether we can identify a discernible effect from the light leak by comparing LMC~X-3 observation 2101010104 (Fig.~\ref{fig:BHXRB_noisepeak_study}a; observation date 2020-08-05) from before the light leak event to observation 6101010108 (Fig.~\ref{fig:BHXRB_noisepeak_study}b; observation date 2023-05-24) from afterwards.
In both cases, a Gaussian is sufficient to describe the noise peak.

Figure~\ref{fig:js_ni6203980141_0mpu7_dark_GTI1_noisepeak_fit} shows the 0.03--10\,keV NICER data of J1727 in the soft state (Obs. ID 6203980141).
We observe a significant flux excess in the 0.3--0.4\,keV overlap range of source and noise peak that cannot be modeled with a Gaussian or t-distribution. Based on the empirical observation that the noise peak looks symmetric in log space, we also attempted fitting a logarithmic t-distribution, which improves the fit only marginally.
We then test whether the fit improves when incorporating a scattering halo, but do not obtain a statistically acceptable fit with physically reasonable parameters.
Given the comparison of J1727 to the well-described data of LMC~X-3 (before and after the light leak) and \eighteentwenty, we conclude that in J1727 there is a soft component present that intrudes the energy range below roughly 0.45\,keV. 
We attempted fitting all NICER data of J1727 with a low energy bound of 0.3\,keV, and observe that this component is persistently seen as a ${\approx}10\%$ uptick in the residuals of many fits across the bright hard state and soft state. 
As we cannot exclude that this component is an extrinsic contamination, we decide to limit the main analysis of this paper to ${>}0.45$\,keV.

\section{Sensitivity of NICER to low accretion disk temperatures}
\label{app:sec:NICER_sensitivity}
\setcounter{figure}{0}

\begin{figure}
    \centering
    \includegraphics[width=0.5\linewidth]{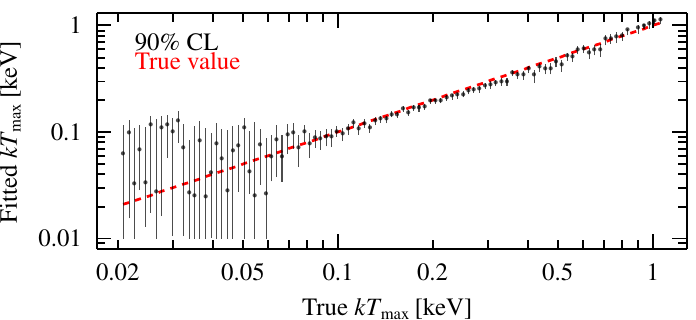}
    \caption{Simulations of a hard state black hole binary with varying disk temperatures. 100 spectra are simulated and fitted in the energy range 0.45--10\,keV. The temperature can be reconstructed well down to around 60\,eV. Below this limit, the parameter values become unconstrained.}
    \label{fig:disk_temperature_limit}
\end{figure}

In this paper, we fit NICER energy spectra from 0.45--10\,keV. Here, we perform simulations to show that we can reliably infer accretion disk temperatures in \swiftj down to a limit of around around 60\,eV.
We create 100 mock spectra using a \texttt{tbabs*simpl*ezdiskbb} model, assuming a constant (absorbed) source flux of $1\times 10^{-9}\,\mathrm{erg}\,\mathrm{cm}^{-2}\,\mathrm{s}$ (0.01--10\,keV), $\Gamma=1.8$, and $f_\mathrm{sc}=0.6$, and 500\,s exposure. These source parameters roughly correspond to \swiftj in the dim hard state. 
We randomize the initial temperature, and rebin and fit the data in the same way as described in Sect.~\ref{sec:data_reduction}.
For a low-energy threshold of 0.45\,keV, we find that we can confidently constrain $kT_\mathrm{max}$ down to around 60\,eV. At this stage, the uncertainties peg at the lower boundary, and other parameters such as $f_\mathrm{sc}$ and $N_\mathrm{ezdiskbb}$ become unconstrained.

\section{Constant interstellar absorption towards \swiftj}
\label{sec:app:abundances}
\setcounter{figure}{0}

\begin{figure}
    \centering
    \includegraphics[width=0.5\linewidth]{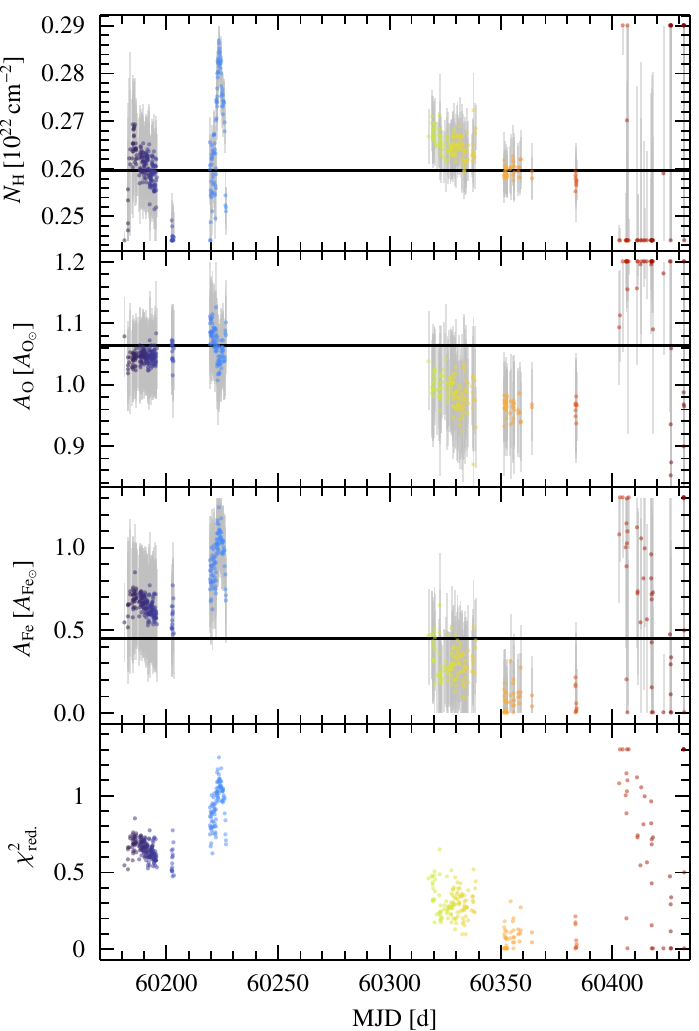}
    \caption{Evolution of the equivalent hydrogen column density $N_\mathrm{H}$, oxygen, and iron abundances in \swiftj. The spectral model is from Eq.~\ref{eq:simplXezdiskbb}. Gray error bars denote 90\% confidence. Horizontal black line denotes the fixed values used in the main analysis of this paper.}
    \label{fig:absorption_estimate}
\end{figure}

For an estimate of the equivalent hydrogen column density $N_\mathrm{H}$ towards J1727, we use the abundances from \citep{Wilms2000a} and cross-sections from \citet{Verner1996a}. 
First, we check whether the assumption of constant absorption throughout the outburst is reasonable (see also, e.g., \citealt{Garcia2019a} for \gx, \citet{Fan2024a} for \eighteentwenty, and \citealt{Chatterjee2024a} for J1727) by letting $N_\mathrm{H}$ free when we fit all orbit-night spectra individually (see Fig.~\ref{fig:absorption_estimate}). 
Across the bright hard state, HIMS, and soft state, the values cluster around a weighted average of $0.26\times 10^{22}\,\mathrm{cm}^{-2}$ with deviations of only a few percent. The dim hard state observations after MJD 60395 peg at the parameter boundaries roughly 10\% away from the weighted average, which we attribute to increased background contamination in these observations. In the transition data (around MJD 60223), clear drifts are visible in all parameters, as well as in the $\chi^2_\mathrm{red.}$ (even though the fit statistic is formally acceptable).

We also allow for the presence of non-solar oxygen ($A_\mathrm{O}$) and iron ($A_\mathrm{Fe}$) abundances (\texttt{tbfeo} in XSPEC terminology). Fits with the \texttt{tbfeo} model are systematically skewed towards unphysically low iron abundances. This bias may be due to features in the NICER spectra arising from systematic uncertainties in the effective area calibration. 
We caution, however, that the derived abundance parameters may be unphysical, and can at present only be interpreted as phenomenological parameters that improve the fit. 

\begin{figure}
    \centering
    \includegraphics[width=0.5\linewidth]{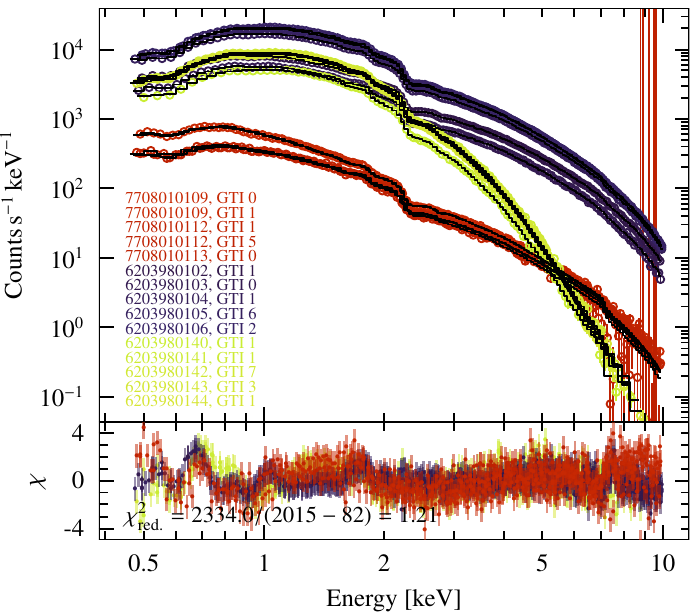}
    \caption{Simultaneous fit of five observations each in the bright hard state (purple), soft state (light green), and dim hard state (red) of \swiftj. A low-energy threshold of 0.45\,keV is used.}
    \label{fig:multifit_all}
\end{figure}

After verifying the assumption of constant absorption, we determine our adopted value of $N_\mathrm{H}$ by simultaneously fitting five bright hard, five soft, and five dim hard state observations with the model \texttt{tbfeo*(ezdiskbb+nthcomp+laor)} (Fig.~\ref{fig:multifit_all}).
The motivation for a simultaneous fit is to minimize systematic uncertainties that may be present at different count rates. In addition, having disk-dominated and corona-dominated spectra in the simultaneous fit minimizes the systematics of the disk/corona model on the absorption.
We obtain $0.2597^{+0.0022}_{-0.0012}\times 10^{22}\,\mathrm{cm}^{-2}$, $A_\mathrm{Fe}=0.45^{+0.09}_{-0.08}$ solar, and $A_\mathrm{O}=1.064^{+0.016}_{-0.022}$ solar.
These values are broadly consistent with the NICER+NuSTAR+IXPE analysis of \citet[][$0.24\pm 0.01\times 10^{22}\,\mathrm{cm}^{-2}$]{Svoboda2024a}, the Insight-HXMT estimates from \citet[][0.10--$0.47\times 10^{22}\,\mathrm{cm}^{-2}$]{Chatterjee2024a}, and the HI4PI survey \citep[][$0.197\times 10^{22}\,\mathrm{cm}^{-2}$]{HI4PI2016a}.

\section{Supplementary material}
\label{app:sec:supplementary}
\setcounter{figure}{0}

\begin{figure}
    \centering
    \includegraphics[width=0.5\linewidth]{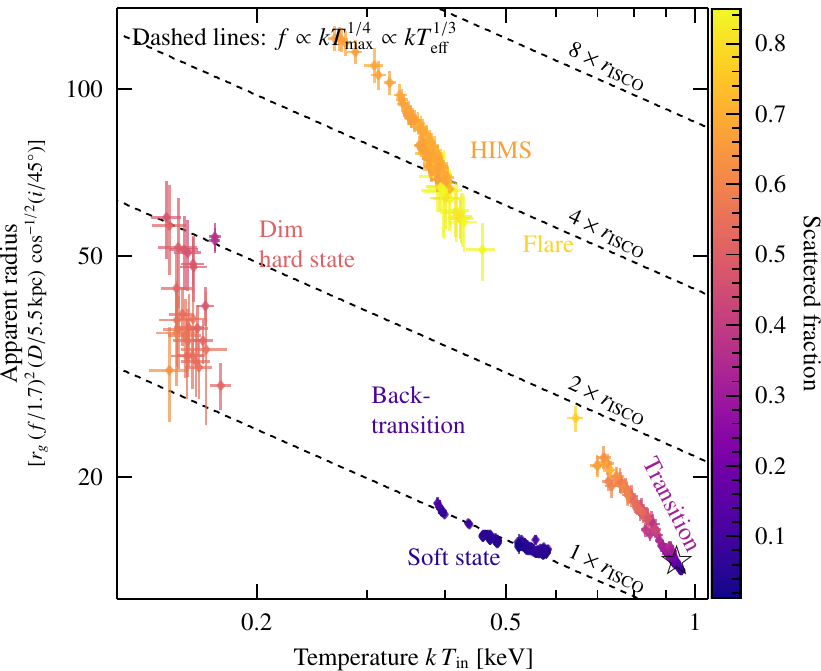}
    \caption{Radius-temperature plane of \swiftj for the model \texttt{simpl\textasteriskcentered diskbb} (similar to Fig.~\ref{fig:ezdiskbb_rin_kt}). The main differences with respect to \texttt{ezdiskbb} is that \texttt{diskbb} does not have a zero-torque boundary condition at the inner edge of the disk, which is more applicable for a truncated disk. The overall hysteresis shape is independent of this parameterization. Orbit-day data is not included in this plot.}
    \label{fig:app:diskbb_rin_kt}
\end{figure}

Appendix Fig.~\ref{fig:app:diskbb_rin_kt} shows fits of the \swiftj orbit-night data with the standard disk model \texttt{diskbb}. Orbit-day data is not included since we primarily want to test whether the overall hysteresis behavior seen in the \texttt{ezdiskbb} fits (Fig.~\ref{fig:ezdiskbb_rin_kt}) is still present when \texttt{diskbb} is used.

\begin{figure}
    \centering
    \includegraphics[width=0.5\linewidth]{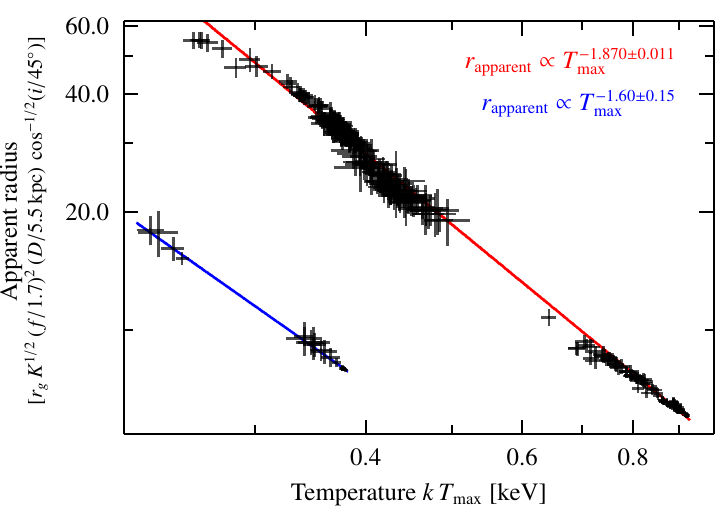}
    \caption{Same data as in the radius-temperature plane of Fig.~\ref{fig:ezdiskbb_rin_kt}. The forward- and backward transition of \swiftj are fitted with a power-law model $r_\mathrm{apparent}=K\cdot T_\mathrm{max}^\gamma$ (red and blue lines, respectively).}
    \label{fig:app:fit_tracks}
\end{figure}

Appendix Fig.~\ref{fig:app:fit_tracks} shows powerlaw fits to the orbit-day and orbit-night data points in the forward-transition (between MJD 60182--60227, denoted as red line) and the backward-transition (MJD 60383.6--60392).

\end{document}